\documentclass[aps,prx,showpacs,twocolumn,superscriptaddress]{revtex4-2}

\usepackage{graphicx}
\usepackage{physics}
\usepackage{wrapfig}
\usepackage{changepage}
\usepackage{amsmath,amssymb,amsthm,mathrsfs,amsfonts,dsfont,mathtools}
\usepackage[caption=false]{subfig}
\usepackage{epsfig}
\usepackage{color,xcolor}
\usepackage{adjustbox}
\usepackage{ bbold }
\usepackage{tabularx}
\usepackage{array}
\usepackage{multirow}
\usepackage{tabu}
\usepackage{bm}
\usepackage{enumerate}
\usepackage{hyperref}
\usepackage{newtxtext,newtxmath}
\usepackage{makecell} %To keep spacing of text in tables
\usepackage{url}
\usepackage{booktabs}
\usepackage{nicefrac} %\biboptions{sort&compress}
\usepackage{makeidx}  % allows for index generation
\usepackage{algorithm}
\usepackage{algpseudocode}
\usepackage{cleveref}
\usepackage{float} 
\usepackage{mwe}

\usepackage{comment}
\usepackage{xurl}
\begin{document}
\title{Improving Quantum Machine Learning via Heat-Bath Algorithmic Cooling}

\author{Nayeli A. Rodr\'{i}guez-Briones}
\email{nayeli.briones@tuwien.ac.at}

\affiliation{
  Atominstitut, Technische Universit{\"a}t Wien, Stadionallee 2, 1020 Vienna, Austria
}
\affiliation{
Miller Institute for Basic Research in Science, University of California Berkeley, CA 94720, USA}

\author{Daniel K. Park}
\email{dkd.park@yonsei.ac.kr}
\affiliation{Department of Statistics and Data Science, Yonsei University, Seoul 03722, Republic of Korea}
\affiliation{Department of Applied Statistics, Yonsei University, Seoul 03722, Republic of Korea}

%%%%%%%%%%%%%%%%%%%%%%%%%%%%%%%%%%%%%%%%%%%%%%%%%%%%%%%%%%%
\begin{abstract}
This work introduces an approach rooted in quantum thermodynamics to enhance sampling efficiency in quantum machine learning (QML). We propose conceptualizing quantum supervised learning as a thermodynamic cooling process. Building on this concept, we develop a quantum refrigerator protocol that enhances sample efficiency during training and prediction without the need for Grover iterations or quantum phase estimation. Inspired by heat-bath algorithmic cooling protocols, our method alternates entropy compression and thermalization steps to decrease the entropy of qubits, increasing polarization towards the dominant bias. This technique minimizes the computational overhead associated with estimating classification scores and gradients, presenting a practical and efficient solution for QML algorithms compatible with noisy intermediate-scale quantum devices.
\end{abstract}

% \keywords{Quantum error mitigation, deep learning, conditional independence, transfer learning}
%%%%%%%%%%%%%%%%%%%%%%%%%%%%%%%%%%%%%%%%%%%%%%%%%%%%%%%%%%%

\maketitle

%%%%%%%%%%%%%%%%%%%%%%%%%%%%%%%%%%%%%%%%%%%%%%%%%%%%%%%%%%%
%%%%%%%%%%%%%%%%%%%%%%%%%%%%%%%%%%%%%%%%%%%%%%%%%%%%%%%%%%%
%%%%%check whether it is easy to read for other people%%%%%

\section{Introduction}
\label{sec:intro}

Quantum machine learning (QML) stands at the intersection of quantum information processing (QIP) and data science, exploring the fundamental limits of physical systems' ability to learn from data and generalize. 
By operating on radically different principles, QML holds the potential of surpassing its classical counterparts in analyzing complex data distributions and identifying intricate patterns.
However, quantum mechanics introduces unique challenges that are absent in classical ML approaches. A critical issue arises from the probabilistic nature of quantum measurements. In QML, both training and inference rely on information extracted from probability distributions associated with the measurements used in the protocol, such as the expectation value of an observable. This process inherently leads to finite sampling errors in QML algorithms.

In theory, adapting Quantum Amplitude Estimation (QAE)~\cite{brassard2002quantum,QSpeedup_MC} to a QML algorithm can quadratically reduce sampling errors. However, QAE requires multiple rounds of Grover-like operations~\cite{suzuki2020amplitude,grinko2021iterative}, which significantly limit its feasibility for Noisy Intermediate-Scale (NISQ) quantum computing. Furthermore, what QAE achieves is often excessive for ML tasks. For instance, in classification problems, it is sufficient to determine the sign of a measured statistic, such as an expectation value, while its magnitude can remain undetermined. This scenario prompts a reevaluation of more practical sampling reduction techniques and motivates investigation into the possibilities of surpassing quadratic improvements in QML.

In response, this work presents an approach rooted in quantum thermodynamics, reframing quantum supervised learning as a thermodynamic cooling process. In this framework, the quantum state represents the input data, and the direction of its population bias (i.e., the sign of its polarization) encodes the model's output. From this viewpoint, we show that reducing the entropy of the quantum states, while preserving their bias direction, mitigates finite sampling errors and improves the quality of statistical estimates. Building on this concept, we develop a quantum refrigerator protocol that enhances sample efficiency in both the training and prediction phases of QML, without relying on complex operations like Grover iterations and quantum phase estimation. Our technique draws inspiration from algorithmic cooling protocols~\cite{boykin2002algorithmic,schulman1999molecular,Park2016,elias2011semioptimal,HBAC_daniel,rodriguez2020novel,oftelie2024dynamic,lin2024thermodynamic,silva2024optimal,PhysRevLett.116.170501}, where entropy is reduced through alternating steps of entropy compression and thermalization. A pivotal element of our approach is the introduction of bidirectional cooling, characterized by distinct fixed points determined by the initial polarization's sign. 
This bidirectional mechanism dynamically reduces entropy through repeated rounds of compression and thermalization, ultimately yielding a state with enhanced polarization in the direction of the initial bias. Remarkably, the protocol operates without requiring prior knowledge of the initial bias direction (i.e. the correct output of the model). Instead, the system is inherently drawn to one of two fixed points, each corresponding to an amplified polarization along the initial bias direction. 
We demonstrate through both theory and numerical analyses that the proposed protocol significantly reduces the number of measurements required to estimate classification scores and gradients, resulting in a substantial reduction in the overall computational cost of QML.
%By reducing the number of measurements needed to estimate classification scores and gradients, this process significantly lowers the computational overhead of QML.

The rest of the paper is organized as follows. Section~\ref{sec:Background} sets up the background and formalizes the classification problem. 
In Section~\ref{sec:QClassifierSingleQubitMeasurement}, we present the framework for variational quantum classifiers, formulating the hypothesis function and classification rule using single-qubit measurements. In Section~\ref{sec:ReductionFiniteSamplingError}, we discuss the challenge of finite sampling errors, presenting bounds on error probabilities for prediction and training. This section also highlights the need for a bidirectional cooling method to enhance the polarization magnitude of the classification score. In Section~\ref{sec:BidirectionalQuantumRefrigerator}, we present the core contribution of this work: the Bidirectional Quantum Refrigerator (BQR), a protocol inspired by Heat-Bath Algorithmic Cooling and tailored for classification tasks in QML. This section begins with a comprehensive analysis of entropy compression, in Section~\ref{sec:Single-ShotEC}. Next, in Section~\ref{sec:BQR_klocal}, we introduce the BQR protocol, supported by theoretical and numerical analyses demonstrating its ability to significantly reduce the number of measurements required to estimate classification scores and gradients. In Section~\ref{sec:BQR_protocol}, we present a practical adaptation of the BQR protocol using $k$-local compression unitaries to enhance its implementability. Finally, Section~\ref{sec:conc} concludes the paper with a discussion of our findings and suggestions for future research directions.
% Unlike previous algorithmic cooling methods, our approach performs cooling rounds with two distinct and attractive fixed points. The fixed point to which the quantum states converge is determined by the unknown initial bias direction of the qubits in the sample. Remarkably, this mechanism functions without the need for prior knowledge of the initial bias direction, as the system is inherently drawn to the corresponding fixed point. The improved qubits enable a reduction in the computational resource overhead associated with the number of repetitions required to estimate the classification score.

\section{Background} \label{sec:Background}
%Classification is a fundamental task in data analysis spanning a wide range of applications. 
Classification is a cornerstone of data analysis, underpinning a wide range of applications. Approaches to this task span from traditional statistical methods and signal processing to advanced techniques in both classical and quantum machine learning. Supervised learning for classification begins with a sample dataset
$
\displaystyle \mathcal{D}=\{ (x_1, y_1), \dots, (x_s, y_s)\} \subseteq \mathbb{R}^d \times \mathbb{Z}_2,
$
where $x_i$ denotes the feature vector, and $y_i$ represents its binary label (or class). Typically, each sample $(x_i,y_i)$ is assumed to be drawn from an unknown joint probability distribution $P_{X,Y}$, i.e., $\displaystyle\left(x_i,y_i\right)\sim P_{X,Y}$. Then, the objective is to construct a hypothesis $f(x)$ such that, for a given loss function $L$, which quantifies the difference between $f(x)$ and the true label $y$, the expected loss $\mathbb{E}_{(x,y)\sim P_{X,Y} }( L(f(x),y))$ is minimized. In practice, the search is often confined to a family of functions parameterized by $\theta$, expressed as  $f(x,\theta)$. Moreover, since the classifier can only be constructed based on the $s$ available samples, the task is typically addressed by minimizing the empirical risk,
$\displaystyle\sum_{i=1}^{s}L\left(f\left(x_i,\theta\right),y_i\right)/s$ with respect to $\theta$. Note that multi-class classifiers with $l>2$ labels can be constructed from binary classifiers using strategies such as One-vs-One or One-vs-Rest comparisons~\cite{GIUNTINI2023109956}.

\section{Quantum classifier with single-qubit measurement} \label{sec:QClassifierSingleQubitMeasurement}

The classification problem can be solved on a quantum computer by training a variational quantum circuit (VQC). In this framework, the canonical form of the hypothesis function is expressed as $f(x,\theta)=\mathrm{sign}\left(\langle x | U^{\dagger}(\theta) M U(\theta) |x\rangle\right)$, where $|x\rangle$ is an $n$-qubit representation of the data, $M$ is a Hermitian matrix and $\mathrm{sign}(\cdot)\in\lbrace -1,+1\rbrace$ denotes the sign function.

Without loss of generality, we can assume that $M$ is an $n$-qubit Pauli operator, i.e., $M\in\lbrace I,X,Y,Z\rbrace^{\otimes n}$. This assumption is well-suited for binary classification, as all Pauli operators have eigenvalues of $+1$ and $-1$.  Consequently, the outcome of any variational quantum binary classifier (VQBC) can be written as $q(x,\theta)=\langle x | U^{\dagger}(\theta) U^{\dagger}_c Z_1 U_cU(\theta) |x\rangle$, where $Z_1$ is the single-qubit Pauli-$Z$ operator acting non-trivially on a single qubit and $U_c$ is an $n$-qubit Clifford operator. The term $q(x,\theta)$ is referred to as the classification score. The VQBC then assigns a label to $x$ based on the sign of this score, i.e., $\displaystyle\mathrm{sign}\left(q\left(x,\theta\right)\right)$. 

An important observation is that any quantum binary classifier can be constructed by training a VQC with a single-qubit Pauli measurement. Moreover, single-qubit measurements are particularly advantageous for mitigating the barren plateaus phenomenon~\cite{10.1038/s41467-018-07090-4,holmes_2022}, which is one of the pressing challenges in variational quantum algorithms. This effectiveness is demonstrated through various techniques, including the use of hierarchical circuit architecture~\cite{grant_hierarchical_2018,pesah2020absence,hur2022quantum,Laurens2023,kim2023classical}, or local observables~\cite{Cost_Cerezo_2021}. In these approaches, $M$ can be selected as $Z_1$, eliminating the need for Clifford transformations.

Now, we rewrite the classification score as
\begin{equation}
\label{eq:score}
    q(x,\theta) = \mathrm{Tr}\left(Z\rho_1(x,\theta)\right),
\end{equation}
where $\rho_1(x,\theta) =\mathrm{Tr}_{n-1}\left( U_cU(\theta) |x\rangle \langle x | U^{\dagger}(\theta) U^{\dagger}_c\right)$ is a single-qubit density operator and $\mathrm{Tr}_{n-1}(\cdot)$ denotes the partial trace over the $n-1$ qubits that are not being measured. Without loss of generality, the Clifford operator $U_c$ can be absorbed into the parameterized unitary.
% Without loss of generality, the Clifford operator $U_c$ can be absorbed into the parameterized unitary, allowing us to express the $n$-qubit final state of the circuit simply as $\rho(x,\theta) = U(\theta) |x\rangle \langle x | U^{\dagger}(\theta)$. As mentioned in the previous paragraph, this scenario arises naturally in many techniques designed to avoid the barren plateau phenomenon. 
By expressing the single-qubit density matrix as 
\begin{equation}
    \rho_1(x,\theta) = \frac{I + \alpha(x,\theta) Z + \beta(x,\theta)X + \gamma(x,\theta) Y}{2},
\end{equation}
the classification rule for an unseen data point $\tilde{x}$ can be written as
% $$
% \tilde{y} =
%     \begin{cases}
%     +1 & \text{if } \alpha(\tilde{x},\theta^{\star}) > 0 \\
%     -1 & \text{if } \alpha(\tilde{x},\theta^{\star}) < 0
%     \end{cases},
% $$
% or equivalently,  
\begin{equation}
\label{eq:rule}
    \tilde{y} = \mathrm{sign}\left(\alpha(\tilde{x},\theta^{\star})\right),
\end{equation}
where $\theta^{\star}$ is the optimal set of parameters obtained by training the VQC. When $\alpha(\tilde{x},\theta^{\star})=0$, the decision can be made at uniformly random.

Since only the $Z$ component of the density matrix affects classification, we assume that the final state of the VQC is
\begin{equation}
    \rho_z(x,\theta) = \frac{I + \alpha(x,\theta) Z}{2}.
\end{equation}
This state can be obtained by adding a parameterized single-qubit gate that rotates the state around the $x$ and $y$ axes of the Bloch sphere.
% , which in turn also changes $\alpha(x,\theta)$ to satisfy $\alpha(x,\theta)^2 + \beta(x,\theta)^2 + \gamma(x,\theta)^2 = \text{constant}$. 
Alternatively, the diagonal state can be prepared by averaging the outcomes of two experiments: $\rho_z(x,\theta) = (\rho_1(x,\theta) + Z\rho_1(x,\theta)Z)/2$.

\section{Reduction of finite sampling error} \label{sec:ReductionFiniteSamplingError}

% The VQBC governed by the classification rule presented in Eq.~(\ref{eq:rule}) necessitates the estimation of $\alpha(x,\theta)$ both during training and when making predictions on unseen data points. Consequently, the same circuit must be executed repeatedly with projective measurements in the Pauli-$Z$ basis to estimate the probabilities $p_0:=\mathrm{Pr}(0) = (1 + \alpha(\tilde{x}, \theta^{\star}))/2$ and $p_1:=\mathrm{Pr}(1) = (1 - \alpha(\tilde{x}, \theta^{\star}))/2$. This represents one of the caveats of QML, introducing an additional computational resource overhead that is absent in classical machine learning.

% The classification score in VQBC is obtained by measuring an expectation value, necessitating the same circuit to be executed repeatedly with projective measurements in the Pauli-$Z$ basis. The finite number of repetitions result in the sampling error. This represents one of the caveats of QML, introducing an additional computational resource overhead that is absent in classical machine learning.

% \subsection{Prediction}
% \label{sec:prediction}
We aim to minimize the sampling error in VQBC, which arises from the finite number of repetitions (or shots) used to estimate the classification score in Eq.~(\ref{eq:score}).
% when predicting the label of an unseen data $\tilde{x}$. 
The relationship between the number of repetitions and the estimation error is described by the Chebyshev inequality 
\begin{equation}
\label{eq:Chebyshev}
    \Pr[|\mu - \langle M\rangle | \ge \epsilon] \le \sigma^2/(k\epsilon^2) 
\end{equation}
where $\mu$ is an average value obtained from $k$ repetitions, $\langle M \rangle$ is the expectation value to be estimated, and $\sigma^2=\langle M^2\rangle - \langle M\rangle^2$ is the variance. 

When predicting the label of unseen data $\tilde{x}$, the classifier uses the sign of the expectation value. Thus, to ensure a correct classification, the estimation error must satisfy $|\mu - \langle M\rangle | < \vert\langle M\rangle\vert $. In our case,  $\langle M\rangle = \alpha(\tilde{x},\theta^{\star})$, and $\sigma^2 = \langle M^2\rangle - \langle M\rangle^2 = 1 - \alpha^2(\tilde{x},\theta^{\star})$. Therefore, the error probability is bounded from above as
\begin{equation}
\label{eq:error_predict}
    \Pr[\mathrm{error}]=\Pr[|\mu - \langle M\rangle | \ge \vert\langle M\rangle\vert] \le \frac{1 - \alpha^2(\tilde{x},\theta^{\star})}{k\alpha^2(\tilde{x},\theta^{\star})}.
\end{equation}

\begin{comment}
Without loss of generality, we can express the error as
\begin{equation}
    \epsilon = \frac{\vert\langle M\rangle\vert^{b}}{c},
    \label{eq:error}
\end{equation}
where a constant $c>1$ and a parameter $b$ determine the desired level of precision. Specifically, $b \ge 1$ if $\vert\langle M\rangle\vert \le 1$ and $b \le 1$ if $\vert\langle M\rangle\vert>1$. For instance, when $\vert\langle M\rangle\vert \le 1$, choosing a larger value for $b$ leads to a decrease in $\epsilon$, imposing a more stringent level of precision. With this formulation, to constrain the error probability to a fixed constant, the required number of repetitions becomes
\begin{equation}
\label{eq:num_shots}
    k\in O\left(\frac{\sigma^2}{\vert\langle M\rangle\vert^{2b}}\right) = O\left(\frac{\langle M^2\rangle - \langle M\rangle^2}{\vert\langle M\rangle\vert^{2b}}\right).
\end{equation}
When $M$ is an $n$-qubit Pauli operator, which is an appropriate choice for VQBC as argued in the previous section, $\vert\langle M\rangle\vert \le 1$ and the above equation simplifies to
\begin{equation}
   k\in O\left(\frac{1}{\vert\langle M\rangle\vert^{2b}}\right),\quad b \ge 1.
\end{equation}

In our case, we wish to estimate $q(x,\theta^{\star}) = \langle M \rangle = \mathrm{Tr}\left(Z\rho_1(x,\theta^{\star})\right) = \alpha(\tilde{x}, \theta^{\star})$. Thus, we have
\begin{equation}
\label{eq:shot}
   k = O\left(\frac{1}{\vert\alpha(\tilde{x}, \theta^{\star})\vert^{2b}}\right),\quad b \ge 1.
\end{equation}
\end{comment}

In contrast, the training process aims to find the set of parameters $\theta$ such that, for all samples $(x_j,y_j)\in\mathcal{D}$, the following conditions hold:
\begin{equation}
   \begin{aligned}
       q(x_j,\theta) > b & \;\text{ if } y_j=+1\\
       q(x_j,\theta) < -b & \;\text{ if } y_j=-1
   \end{aligned}
\end{equation}
where $0\le b <1$ defines the margin. These conditions can be expressed as a single inequality that the training samples should satisfy: $y_jq(x_j,\theta) - b > 0 \;\forall j=1,\ldots,s$.
%\begin{equation}
 %   y_jq(x_j,\theta) - b > 0 \;\forall j=1,\ldots,s.
%\end{equation}
Thus, the training process seeks to minimize the hinge loss, defined as $l(\theta|x_j,y_j) = \max(0, b-y_jq(x_j,\theta))$. The gradient of the hinge loss is given by
\begin{equation}
\label{eq:grad}
    \frac{dl}{d\theta} = \begin{cases}
        -y_j\frac{\partial q(x_j,\theta)}{\partial \theta} & \text{ if } y_jq(x_j,\theta) < b \\
        0 & \text{otherwise}
    \end{cases}.
\end{equation}
Equation~(\ref{eq:grad}) highlights that implementing gradient-based optimization requires estimating $q(x_j,\theta)$ with an error smaller than $\vert q(x_j,\theta) - b\vert$ to accurately assess whether $yq(x_j,\theta)<b$. Consequently, the probability of making an incorrect decision at this step is bounded by
\begin{equation}
\label{eq:error_train}
    \Pr[\mathrm{error}]\le \frac{1 -  \alpha^2(x_j,\theta)}{k(\alpha(x_j,\theta)-b)^2}.
\end{equation}
Additionally, the estimation error must be sufficiently small to ensure effective parameter updates. The gradient's sign indicates the direction of the parameter update, while its magnitude determines the step size in that direction. Notably, correctly estimating the sign of the gradient is more critical than its magnitude, as the latter can be scaled by an adaptively adjusted learning rate. This implies that $\partial q(x_j,\theta)/\partial \theta$ must be estimated within the error margin smaller than $\vert \partial q(x_j,\theta)/\partial\theta\vert$. This requirement results in an upper bound on the error probability similar to Eq.~(\ref{eq:error_predict}), as shown in Appendix~\ref{sec:appendix_C}.
% Combining these considerations, the estimation error should satisfy $\vert \mu - \langle M\rangle \vert < \min( \vert q(x_j,\theta)\vert, \vert q(x_j,\theta) - b\vert)$. 
% Thus, the error probability is bounded from above as
% \begin{equation}
% \label{eq:error_train}
%     \Pr[\mathrm{error}]\le \frac{1 -  \alpha^2(x_j,\theta)}{k\min\lbrack\alpha^2(x_j,\theta),(\alpha(x_j,\theta)-b)^2\rbrack}.
% \end{equation}

The error probabilities in Eqs.~(\ref{eq:error_predict}) and~(\ref{eq:error_train}) highlight the importance of having a large magnitude for $\alpha(x, \theta)$ to minimize the number of repetitions required. This observation motivates the development of a protocol that increases $\vert\alpha(\tilde{x}, \theta)\vert$. Such a protocol cannot be a locally unitary process on the single qubit, as it must alter the purity of its state. Various algorithmic cooling techniques, based on engineering the system-bath interaction, exist that can increase the population of a chosen basis state~\cite{boykin2002algorithmic,schulman1999molecular,Park2016,elias2011semioptimal,lin2024thermodynamic,HBAC_daniel,PhysRevLett.116.170501,1367-2630-19-11-113047,PhysRevLett.119.050502,alhambra2019heat,taranto2024efficiently,silva2024optimal}. 
% For all input states, these techniques can only increase either $p_0$ or $p_1$ predetermined beforehand. 
However, it is crucial to note that the sign of $\alpha(x, \theta)$ is not known a priori in the QML context. Thus, conventional algorithmic cooling techniques, which solely increase the population of a predetermined basis state, cannot achieve our goal. Instead, we require a bidirectional protocol capable of dynamically transforming the single-qubit density matrix as follows:
\begin{equation}
\label{eq:goal}
\frac{I + \alpha Z}{2} \rightarrow \frac{I + \alpha' Z}{2}\text{ with } 
 \begin{cases}
\alpha' > \alpha & \text{ if } \alpha > 0 \\
\alpha' < \alpha & \text{ if } \alpha < 0
\end{cases}.
\end{equation}

Bidirectional cooling can also be motivated from the perspective of quantum state discrimination. To illustrate this, consider splitting the sample dataset as $\mathcal{D} = \mathcal{D}_+ \cup \mathcal{D}_-$ where $\displaystyle\mathcal{D}_{\pm}=\lbrace(x_i^{\pm},\pm1)\rbrace_{i=1}^{s_{\pm}}$ denotes the subset containing only data labeled as $\pm 1$, and $s_{\pm}$ denotes the size of the $\pm 1$-class dataset. 
% The optimization process should find the set of parameters such that $\rho_z(x_i^{+},\theta^{\star}) = (I+|\alpha(x_i^{+},\theta)|Z)/2$ $\forall x_i^{+} \in \mathcal{D}_+$ and $\rho_z(x_i^{-},\theta^{\star}) = (I-|\alpha(x_i^{-},\theta)|Z)/2$ $\forall x_i^{-} \in \mathcal{D}_-$. 
As noted in Ref.~\cite{PhysRevA.110.022411}, the optimization process can be interpreted as finding $\theta$ that maximizes the discrimination between the density matrices $\displaystyle\rho_+=\sum_{i=1}^{s_{+}}\rho_z(x_i^{+},\theta)/s_{+}$ and $\displaystyle\rho_{-}=\sum_{i=1}^{s_{-}}\rho_z(x_i^{-},\theta)/s_{-}$, which represent the ensemble of data points in each class, with a high success probability. In this picture, the empirical risk (i.e. training loss) is bounded from below by the error probability in distinguishing the two ensembles. These ensembles can also be expressed as $\displaystyle\rho_{\pm}=(I+\bar{\alpha}_{\pm}Z)/2$, where $\bar{\alpha}_{\pm}$ is the average polarization of $\rho_z(x_i^{\pm},\theta)$. Assuming $s_+ = s_{-}$, the error probability of discriminating $\rho_+$ and $\rho_{-}$ is 
\begin{equation}
\label{eq:qsd_success}
    \begin{aligned}
\Pr[\mathrm{error}] & = \frac{1}{2}\left(1 - \left\Vert \frac{s_+}{s}\rho_+-\frac{s_-}{s}\rho_-\right\Vert_1\right)\\
   % & = \frac{1}{2}- \frac{1}{4}\left\Vert \frac{\left(\bar{\alpha}_+ -\bar{\alpha}_{-}\right)Z}{2}\right\Vert_1 \\
   & = \frac{1}{2} - \frac{1}{4}\vert\bar{\alpha}_+-\bar{\alpha}_{-}\vert .
   \end{aligned}
\end{equation}
Therefore, increasing the magnitude of the polarizations via bidirectional cooling reduces the lower bound of the empirical risk, improving the classifier's performance.

While the main results of this work are demonstrated through VQBC, the transformation outlined in Eq.~(\ref{eq:goal}) is applicable to any binary classifier established based on the classification score given in Eq.~(\ref{eq:score}). An example of this broader applicability is provided in Appendix~\ref{sec:appendix_D}.

In the following sections, we introduce an approach rooted in quantum thermodynamics to address the challenge of finite sampling error. Specifically, this approach achieves the transformation in Eq.~(\ref{eq:goal}) without requiring any prior information about $\alpha$.
%The refrigerator uses a number of copies of the density matrix. But the cooling is efficient that the overall number of repeating the VQC is smaller than when not using the SDQF.

\section{Bidirectional Quantum refrigerator}\label{sec:BidirectionalQuantumRefrigerator}

Heat-Bath Algorithmic Cooling (HBAC) methods have demonstrated the ability to cool down quantum systems by increasing the ground state population $p_0$ via entropy compression unitaries and thermalization steps~\cite{boykin2002algorithmic,schulman1999molecular,Park2016,elias2011semioptimal,lin2024thermodynamic,HBAC_daniel}. 
However, HBAC protocols have been specifically optimized to increase $p_0$, i.e., resulting in a unidirectional change in $\alpha(x,\theta)$. Moreover, these protocols rely on prior knowledge of the sign of $\alpha(x,\theta)$ to design optimal compression unitaries that ensure effective cooling. These characteristics render HBAC protocols unsuitable for classification problems, where the sign of $\alpha(x,\theta)$---which determines the gradient direction or predicts the label---must remain unknown a priori.

In this study, we extend the foundational elements of HBAC to classification problems for the first time. We introduce a novel quantum refrigerator designed to reliably increase polarization in the bias direction, regardless of the unknown parameter $\alpha(x,\theta)$. Specifically, we design and utilize entropy compression gates that obey the transformation given by Eq.~(\ref{eq:goal}) for unknown polarization, functioning effectively in both bias directions. Based on these findings, we introduce a new family of cooling protocols called Bidirectional Quantum Refrigerators (BQRs).  Moreover, in contrast to conventional HBAC methods, our protocols operate cyclically---recycling output systems, after the removal of the enhanced qubits---to generate multiple subsequent enhanced qubits, thereby reducing the required qubit resources.

\begin{comment}
We discovered that a particular example that satisfies this condition is given by the entropy compression of the minimal 3-qubit HBAC scenario, which is a special case of HBAC in which the optimal circuit consists of the iteration of identical rounds~\cite{Park2016,PhysRevLett.116.170501,baugh2005experimental}. 
In this minimal 3-qubit setup, the entropy compression always swaps the populations of $\ket{011}$ and $\ket{100}$ states---see Fig.\ref{fig:QRefrigerator} for the explicit gate representation, denoted by $U_{\textsc{c}_3}$. However, for $n>3$ the rounds of HBAC are optimized assuming the state of the system is known and fixing the bias direction that needs to be increased. In this work, we generalize the property of the 3-qubit HBAC to circuits with $n>3$, increasing the population of the target qubit in the direction of the unknown bias of the thermal reservoir. 
\end{comment}

%%%%%

\subsection{Single-Shot Entropy Compression}\label{sec:Single-ShotEC}
%\textit{Algorithmic Cooling: Single-Shot Entropy Compression.---}

As a foundational step, we present results for single-shot entropy compression, a process that enhances qubit polarization through unitary operations. This method reduces the entropy of a target subsystem by implementing reversible transformations that redistribute the internal entropy across the entire system~\cite{Sorensen1989, Sorensen1990, Sorensen1991}. Commonly referred to as AC (Algorithmic Cooling) entropy compression, this approach serves as a building block. By selecting transformations that exhibit bidirectional cooling, we extend its principles to design the quantum refrigerator introduced in subsequent sections, where entropy compression is performed iteratively over multiple rounds.

The system consists of $n$ qubits: one target qubit and $n-1$ auxiliary qubits that assist in entropy compression. The goal is to improve the polarization of the target qubit through unitary operations. In general, the explicit form of the optimal entropy compression depends on the state of the system. It is established that, to increase the ground-state population of a target qubit, the effect of the optimal compression is to diagonalize the state and reorder the diagonal elements of the system's total density matrix in descending order. %~\cite{Park2016,rodriguez2020novel,oftelie2024dynamic} 
However, no analysis has been conducted on achieving a bidirectional effect for an unknown $\alpha$.

We found that, when the system consists of $n$ identical qubits, applying the same optimal compression for $\alpha>0$ to a scenario where the sign of $\alpha$ is unknown, achieves the transformation described in Eq.~(\ref{eq:goal}). This result is stated formally in the following theorem.\\

\textbf{Theorem 1. (Optimal Bidirectional Single-Shot Entropy Compression on Identical Qubits).} 
\textit{Consider $n$ identical qubits with an unknown population parameter $\alpha$, each qubit in the state $\rho = \begin{pmatrix} p & 0 \\ 0 & 1-p \end{pmatrix}$, where $\displaystyle p=\frac{1+\alpha}{2}$. The total state $\rho^{\otimes n}$ is a diagonal density matrix, with all elements expressed as $p^{n-j}(1-p)^j$ for $j = 0, 1, \dots, n$.
%(multiple entries may share the same value/label). 
The optimal bidirectional single-shot compression consists of reordering the populations following the order: $p^n, p^{n-1}(1-p), p^{n-2}(1-p)^2, \dots, (1-p)^n$.}\\

\textit{Proof.---} 
For $p>1/2$ ($\alpha > 0$), the optimal entropy compression over all possible global unitaries reorders the elements in decreasing order as:
$$
p^n > p^{n-1}(1-p) > p^{n-2}(1-p)^2 > \dots > (1-p)^n,\; {\rm if}\; p>1/2.
$$
This reordering optimally maximizes the population of the $\ket{0}$ state  of the target qubit, thereby enhancing $\alpha$. This result has been shown using majorization arguments in Ref.~\cite{rodriguez2020novel} and techniques for computing ergotropy in Ref.\cite{oftelie2024dynamic}.
On the other hand, if $p \leq 1/2$  ($\alpha \leq 0$), implementing the same reordering results in an increasing order:
$$
p^n \leq p^{n-1}(1-p) \leq p^{n-2}(1-p)^2 \leq \dots \leq (1-p)^n,\; {\rm if}\; p\leq 1/2
$$
as $p \leq 1-p$. By majorization, this ordering optimally enhances the population of the $\ket{1}$ state  of the target qubit through unitary operations, thereby reducing $\alpha$.
Thus, the optimal single-shot entropy compression for $n$ identical qubits achieves an optimal bidirectional cooling transformation described in Eq.~(\ref{eq:goal}).~$\blacksquare$

\begin{figure*}[htbp]
\centering
\includegraphics[width=1\linewidth]{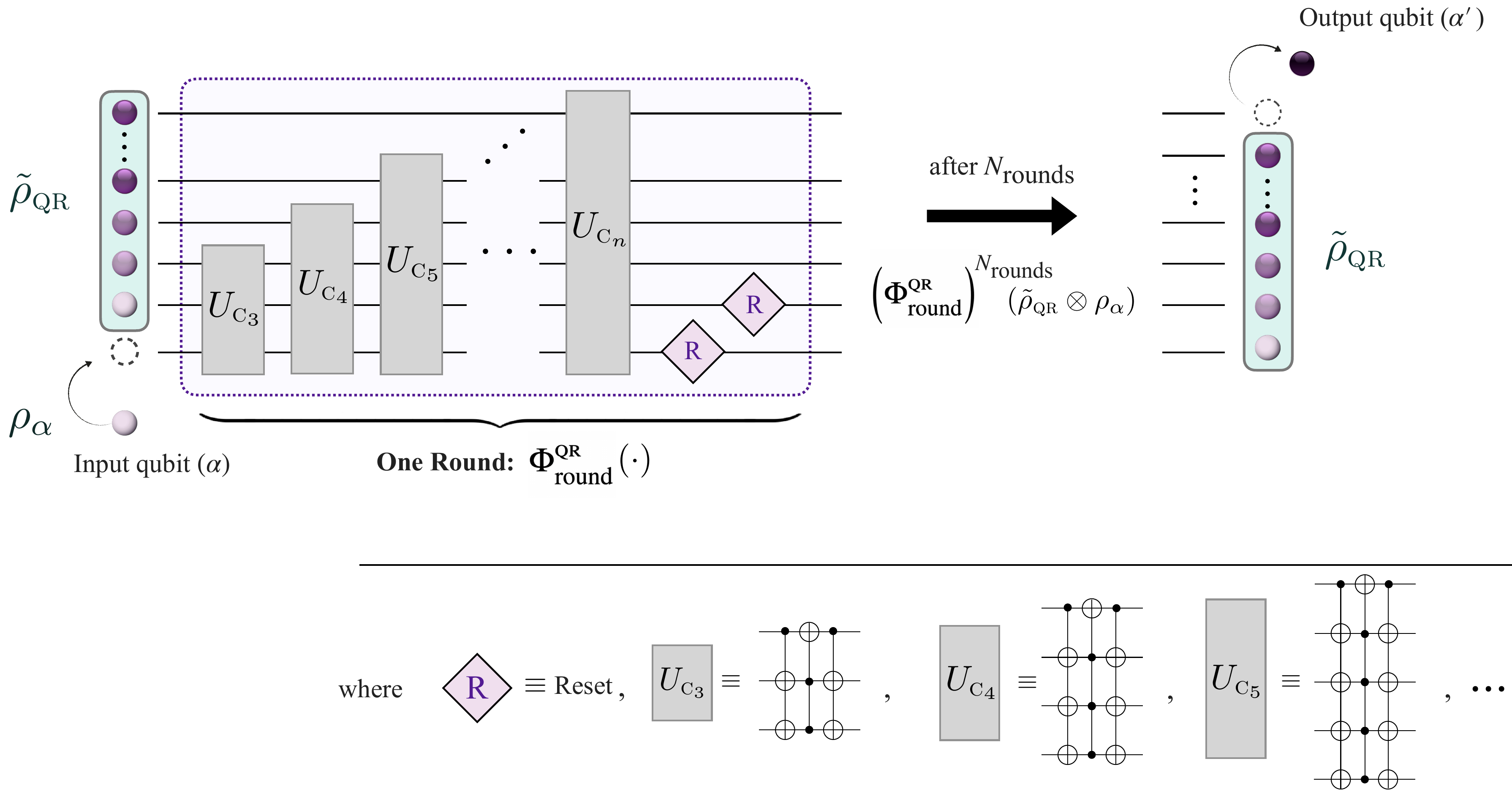}
\caption{Bidirectional quantum refrigerator protocol circuit operating on $n$ qubits, including $m$ reset qubits ($m=2$ in this figure).}
\label{fig:QRefrigerator}
\end{figure*}

The explicit gate representation of a unitary operation that achieves the optimal reordering can be designed based on $n$. For example, for $n=3$,  that entropy compression requires swapping the populations of $\ket{011}$ and $\ket{100}$ states, and its explicit representation is shown in Fig.~\ref{fig:QRefrigerator}, denoted by $U_{\textsc{c}_3}$.

\begin{figure}[t]
\centering
\includegraphics[width=1\linewidth]{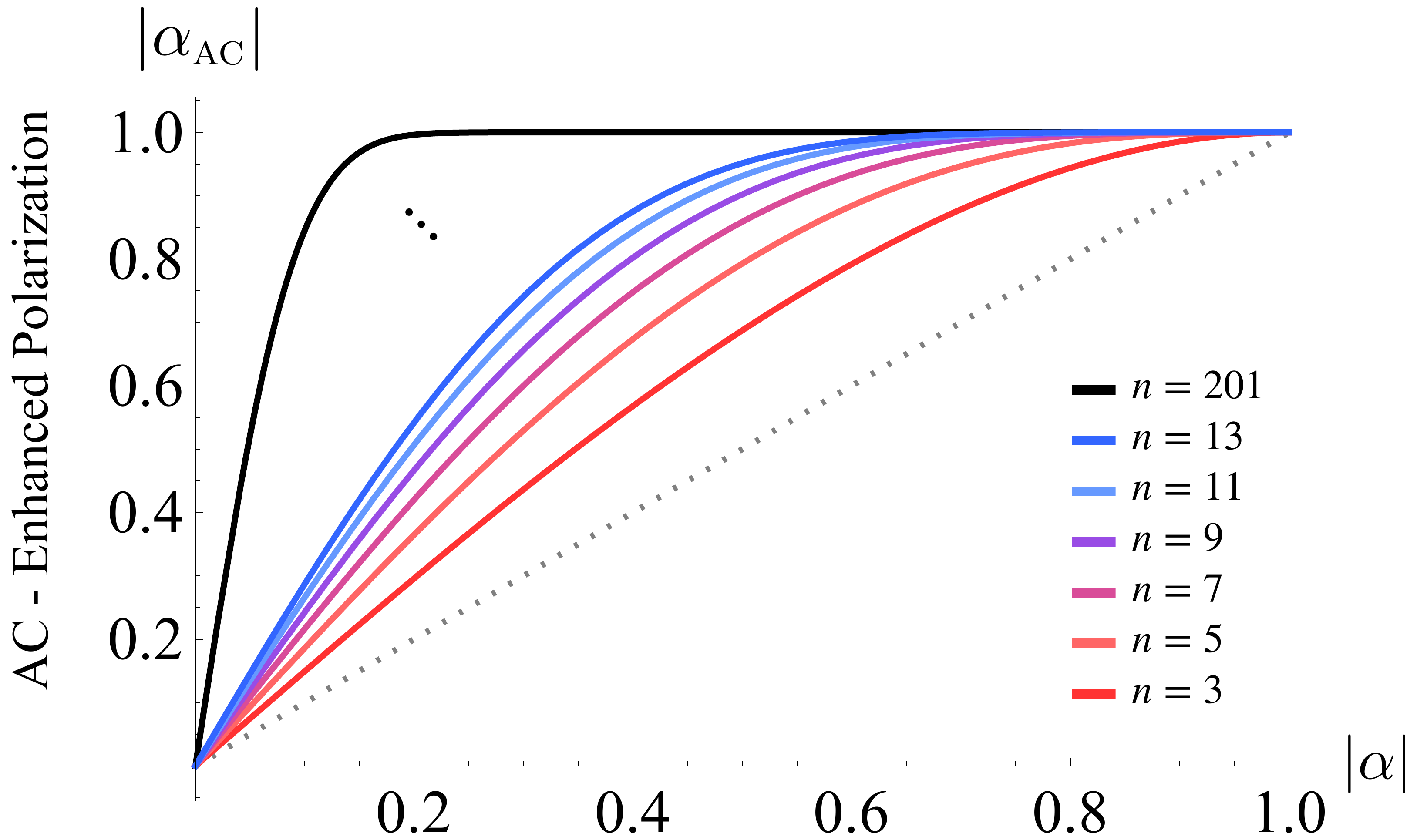}
\caption{Enhanced polarization $\alpha_\textsc{ac}$
  resulting from optimal single-shot entropy compression, plotted as a function of the initial 
$|\alpha|$ for different numbers of qubits. The black dotted line represents the baseline, initial polarization.}
\label{fig_single_shot_polarization}
\end{figure}

The optimal entropy compression on $n$ qubits with an initial polarization of $\alpha$ enhances the polarization of the target qubit as follows~\cite{rodriguez2020novel,oftelie2024dynamic,lin2024thermodynamic}:
\begin{equation}
\alpha \to \alpha_\textsc{ac} = 2\left[\sum^{\lfloor(n-1)/2\rfloor}_{i=0}\binom{n}{i}p^{n-i}\left(1-p\right)^{i}\right]-1,
\label{eq:alphaAC}
\end{equation}
where $p=(1+\alpha)/2$, and $\lfloor\cdot\rfloor$ denotes the floor function. %This improvement corresponds to odd $n$ and takes the same value for $n+1$. 
Figure~\ref{fig_single_shot_polarization} shows the $\alpha_\textsc{ac}$ for different $n$.
%
%In the thermodynamic limit, 
This maximum achievable polarization can be approximated as 
\begin{equation}
\displaystyle
    \alpha_\textsc{ac}\simeq {\rm erf}(\xi):=\frac{2}{\sqrt{\pi}}\int^\xi_0 e^{-t^2}dt
\label{eq:alphaACapprox}
\end{equation}
where $\displaystyle\xi =n\alpha/\sqrt{2n(1-\alpha^2)}$~\cite{rodriguez2020novel}. %In the low polarization regime, $\xi\approx\alpha\sqrt{n/2}$, and $\displaystyle\alpha_\textsc{ac}\approx \alpha\sqrt{2n/\pi}$~\cite{oftelie2024dynamic}.

\begin{figure}[t]
\centering
\includegraphics[width=1\linewidth]{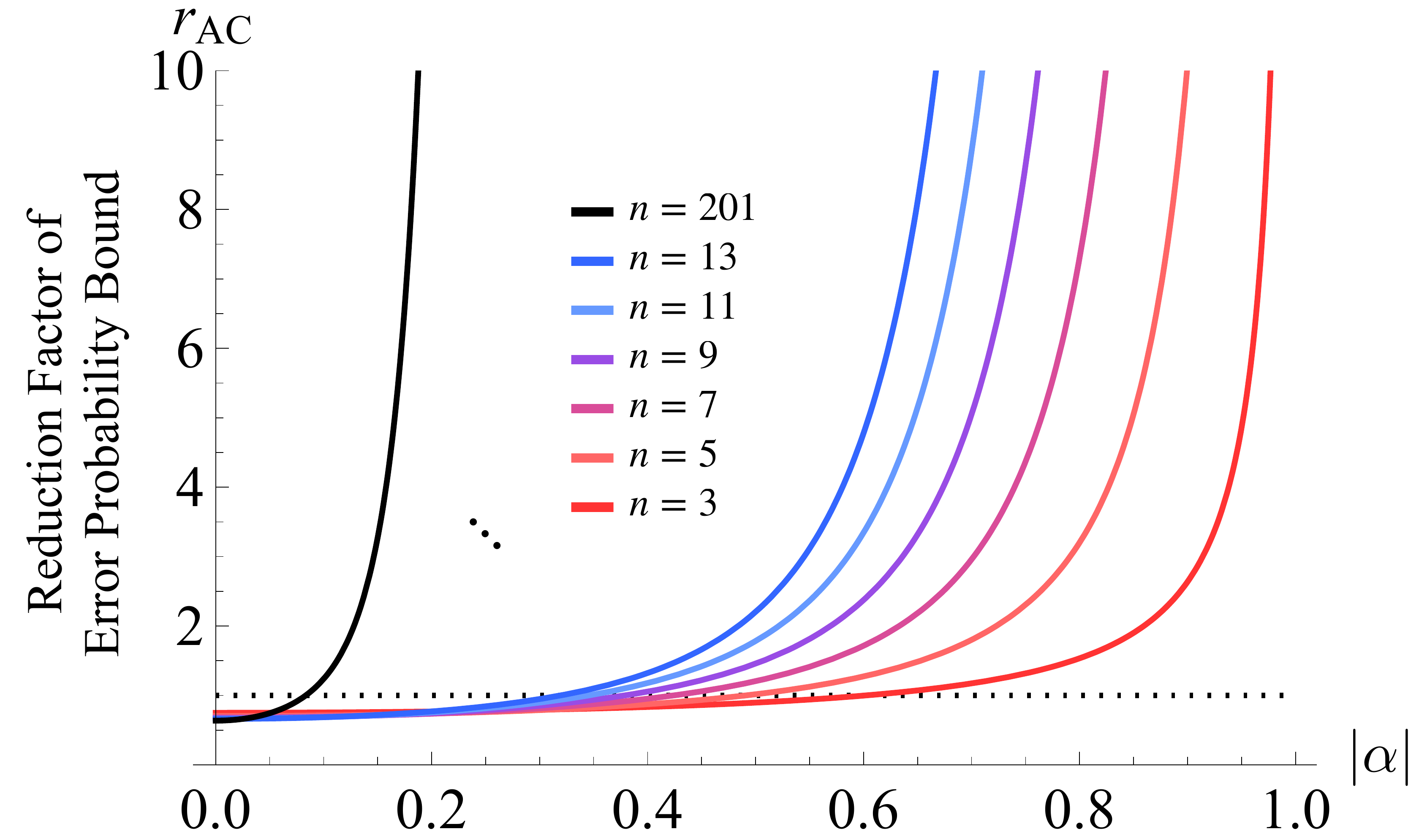}
\caption{Reduction factor of the error probability bound after enhancing polarization via AC entropy compression, as a function of the initial polarization $|\alpha|$. The results are shown for different values of $n$, using the error probability bound given by Eq.~(\ref{eq:error_predict}).}
\label{fig_single_shot_reduced_resources}
\end{figure}

After enhancing the polarization via $n$-qubit optimal entropy compression, the sampling error probability bound,  Eq.~(\ref{eq:error_predict}), is reduced by a factor of
\begin{equation}
\displaystyle r_\textsc{ac}=
   %\frac{{\rm Pr[error]}}{{\rm Pr}_\textsc{ac}{\rm [error]}}=
   \frac{1-\alpha^2}{1-\alpha_\textsc{ac}^2}\left(\frac{\epsilon\left(\alpha_\textsc{ac}\right)}{\epsilon\left(\alpha\right)}\right)^2\frac{1}{n}
\label{eq:kkcom}
\end{equation}
where $\epsilon(\alpha)$ represents the requested precision. This result assumes that the same amount of qubit resources is used, i.e. $k$ repetitions with the original qubits versus $k/n$ repetitions with the enhanced qubits. Figure~\ref{fig_single_shot_reduced_resources} shows the reduction factor as a function of the initial polarization $\alpha$ for different values of $n$.

Based on the error probability bound given in Eq.~(\ref{eq:error_predict}) and the enhanced polarization approximation, $\displaystyle r_\textsc{ac}=\frac{1}{n}\frac{1-\alpha^{-2}}{1-{\rm erf}(\xi)^{-2}}$. 
The behavior of $r_\textsc{ac}$ in the different polarization regimes is as follows: \textbf{(1)} In the low polarization regime,  $\alpha_\textsc{ac}\approx \alpha \sqrt{2n/\pi}$, resulting in $r_\textsc{ac}\approx\frac{2(1-\alpha^2)}{\pi-2n\alpha^2}$.
%
%Thus, for $|\alpha|\approx0$, $r_\textsc{ac}\to \sim 2/\pi$. 
\textbf{(2)} In the intermediate polarization regime, $r_\textsc{ac}$ grows exponentially as $\propto e^{x}$, where $x\sim\frac{n\alpha^2}{2(1-\alpha^2)}$. \textbf{(3)} For high polarization regime, in the limit of $\alpha\to1$, $r_\textsc{ac}$ diverges as the denominator approaches zero. Further details are provided in Appendix~\ref{sec:appendix_A}.
The area under the curves in Fig.~\ref{fig_single_shot_reduced_resources}  diverges, and the rate of divergence grows as $n$ increases. This indicates that, although the compression strategy does not provide an advantage in the low polarization regime, the sampling error probability bound improves for all $n\geq3$ on average when the polarization is uniformly distributed, with greater improvement as $n$ increases. Furthermore, the extent of the low-polarization region can be effectively reduced by increasing $n$.
% The sampling error probability bound is, on average, enhanced for all $n\geq3$ on average, with greater improvement as the value of $n$ increases. Note that, although locally, in the low polarization regime, the compression strategy does not provide an advantage on average across all polarizations, the enhancement is greater than 1 and increases with the number of qubits. Furthermore, it is possible to reduce the low polarization region where there is no local advantage by increasing the value of $n$.

%
%

%\begin{figure}[h]
%\centering
%\includegraphics[width=1\linewidth]{Logkkcromn.pdf}
%\caption{Log version of Fig.~\ref{fig_single_shot_reduced_resources}}
%\label{fig_log_single_shot_reduced_resources}
%\end{figure}

Now, to go beyond the single-shot entropy compression, we present a novel method for enhancing the polarization of qubits. This approach, inspired by HBAC, includes rounds that allow refreshing qubits to continue compressing entropy. We design explicit circuits that  circumvent the limitations that render conventional HBAC ineffective in situations where the polarization can be either negative or positive, and where both the value and sign are unknown. Moreover, our approach accounts for the need to repeat the process for each enhanced qubit for sampling. Thus, we adjusted the method to recycle part of the compressed system and use it as a refrigerator, as detailed in the following section.

%%%%%%%%%%%%%%%%%%%%%%%%%%%%%%%%%%%%
%%%%%%%%%%%%%%%%%%%%%%%%%%%%%%%%%%%%
%

\subsection{Bidirectional Quantum Refrigerator}\label{sec:BQR_protocol}
%\textit{Bidirectional Quantum Refrigerator.---} 
The system setup consists of a string of $n$ qubits: one target qubit to be improved, $m$ reset qubits that can be replaced by fresh qubits from the sample, and $n-m-1$ auxiliary qubits that assist in the entropy compressions.
The BQR purifies a target qubit through $N_{\rm rounds}$, with each round consisting of two parts: first, a set of unitaries that increase the dominant bias of the target qubit, followed by the replacement of $m$ reset qubits, to pump the excess entropy out of the system.
The protocol with the explicit form of the entropy compression unitaries is shown in Fig.~\ref{fig:QRefrigerator}. In each round, the set of unitaries $Uc_3$,...,$Uc_n$ is implemented sequentially in a stair-like manner, as depicted in Fig.~\ref{fig:QRefrigerator}, where each $U_{\textsc{c}_j}$ swaps the states $|0\rangle |1\rangle^{\otimes(j-1)}$ and $|1\rangle|0\rangle^{\otimes (j-1)}$. The global effect of the set of unitaries is given by 
\begin{align*}
    U_\textsc{qr}(n)&=U_{\textsc{c}_n}(\mathds{1}_{2}\otimes U_{\textsc{c}_{(n-1)}})(\mathds{1}_{2^2}\otimes U_{\textsc{c}_{(n-2)}})...(\mathds{1}_{2^{(n-3)}}\otimes U_{\textsc{c}_3}),
\end{align*}
This can be represented as the following block matrix
\begin{align}
U_\textsc{qr}(n)=\begin{bmatrix}
\mathds{1}_3 & 0 &0\\
0 & \sigma_\textsc{qr}(n) & 0\\
0& 0 & \mathds{1}_3 
\end{bmatrix},
\end{align}
where $\mathds{1}_{3}$ is the identity matrix with a diagonal of size three, $\sigma_\textsc{qr}(3)=\sigma_x$, and
\begin{align}
        \sigma_\textsc{qr}(n):=\begin{bmatrix}
        \mathds{1}_2\otimes\begin{bmatrix} 
        \sigma_\textsc{qr}(n-1) &0\\
        0&\mathds{1}_2
        \end{bmatrix} &0\\
        0&\sigma_x
    \end{bmatrix},\;\;\;{\rm for} \;\;\; n>3.
\end{align}
The effect of a single round on a system of $n$ qubits in the state $\rho$ is given by
\begin{equation}
    \rho\to \Phi^{\textsc{qr}}_{\rm round}(\rho):={\rm Tr_{m}}\left[ U_\textsc{qr}(n)\rho U^\dagger_\textsc{qr}(n)\right]\otimes\rho_\alpha^{\otimes m}.
\end{equation}

Note that $U_{\textsc{c}_3}$ corresponds to the optimal AC entropy compression for $n=3$ and is also iterated in the optimal 3-qubit HBAC. In fact, the 3-qubit HBAC represents a special case where the optimization of the unitary in all rounds consists of the iteration of identical rounds~\cite{Park2016,PhysRevLett.116.170501,baugh2005experimental}. This round is also a specific instance within the rounds of our BQR family.
On the other hand, the unitaries $U_{\textsc{c}(n>3)}$ are not optimal compressions for $n>3$ but satisfy bidirectional cooling.

\begin{comment}
We discovered that a particular example that satisfies this condition is given by the entropy compression of the minimal 3-qubit HBAC scenario, which is a special case of HBAC in which the optimal circuit consists of the iteration of identical rounds. 
In this minimal 3-qubit setup, the entropy compression always swaps the populations of $\ket{011}$ and $\ket{100}$ states---see Fig.\ref{fig:QRefrigerator} for the explicit gate representation, denoted by $U_{\textsc{c}_3}$. However, for $n>3$ the rounds of HBAC are optimized assuming the state of the system is known and fixing the bias direction that needs to be increased. In this work, we generalize the property of the 3-qubit HBAC to circuits with $n>3$, increasing the population of the target qubit in the direction of the unknown bias of the thermal reservoir. 
\end{comment}

In contrast to HBAC, where the system's initial state is a product state of $n$ fresh qubits, our BQR reuses the output system for subsequent preparations.
Namely, after preparing an enhanced qubit, the remaining $n-1$ qubits are recycled as the updated system for the quantum refrigerator, along with a new fresh qubit to complete the $n$ qubits needed for the circuit to run again. 
The recycled set of qubits reaches an output steady-state $\tilde{\rho}_\textsc{qr}(n,\alpha,N_{\rm rounds},m)$, which reduces the number of resources required to prepare enhanced qubits ($\displaystyle\sim m N_{\rm rounds}+1$, instead of $\displaystyle n+mN_{\rm rounds}$), and improves the convergence rate. %The fixed point of the body of the refrigerator depends on the number of qubits $n$, the initial polarization $\alpha$, and the number of rounds $N_{\rm rounds}$. 
\begin{figure}[t]
\centering
\includegraphics[width=0.97\linewidth]{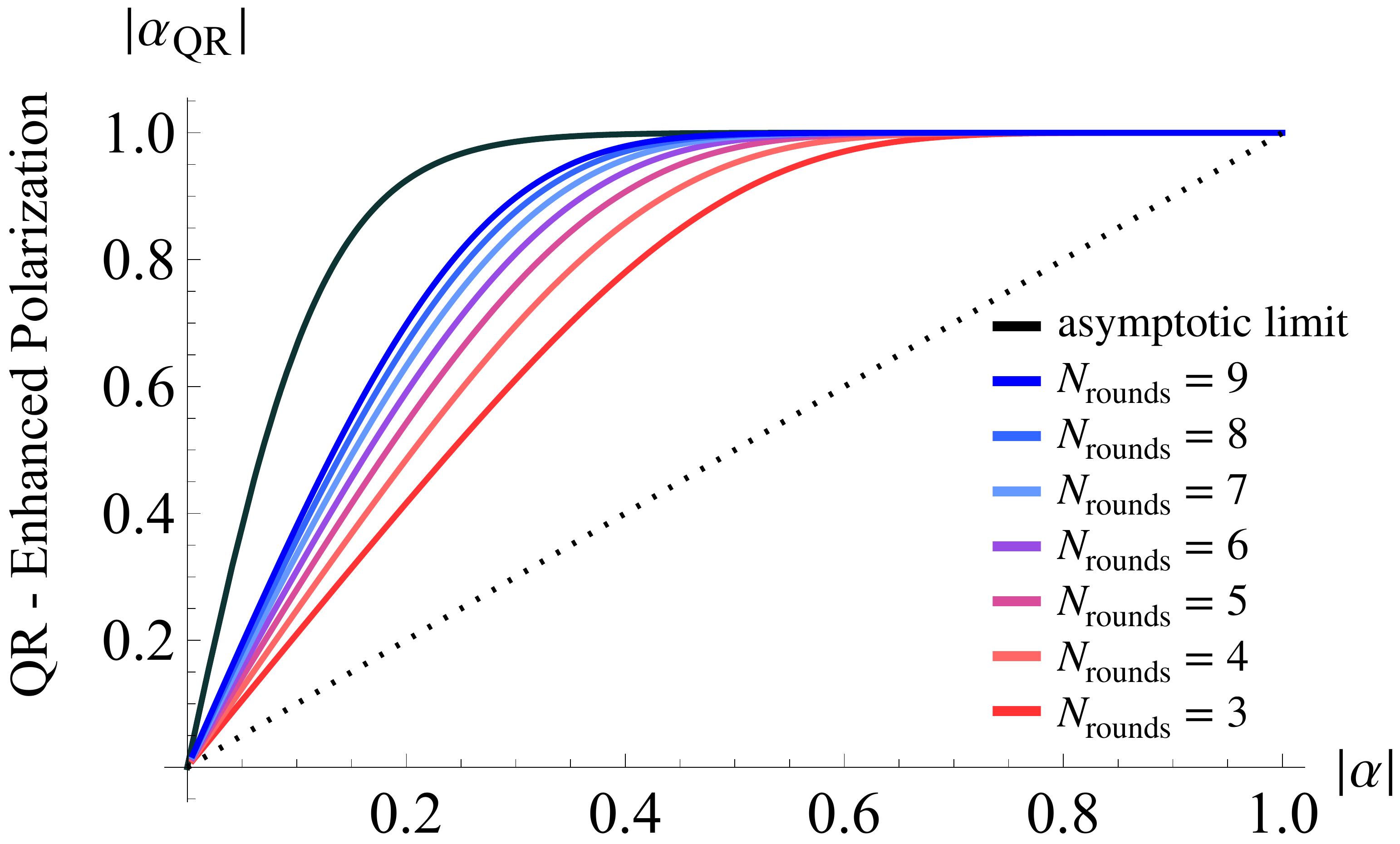}
\caption{Enhanced polarization $\alpha_\textsc{qr}$
 for the quantum refrigerator operating with $n=5$ as a function of the initial 
$|\alpha|$ for different number of rounds. The black dotted line represents the baseline for the initial polarization, and the solid black line corresponds to the asymptotic polarization.}
\label{fig:alphaQRn5}
\end{figure}

\begin{figure}[t]
\centering
\includegraphics[width=1\linewidth]{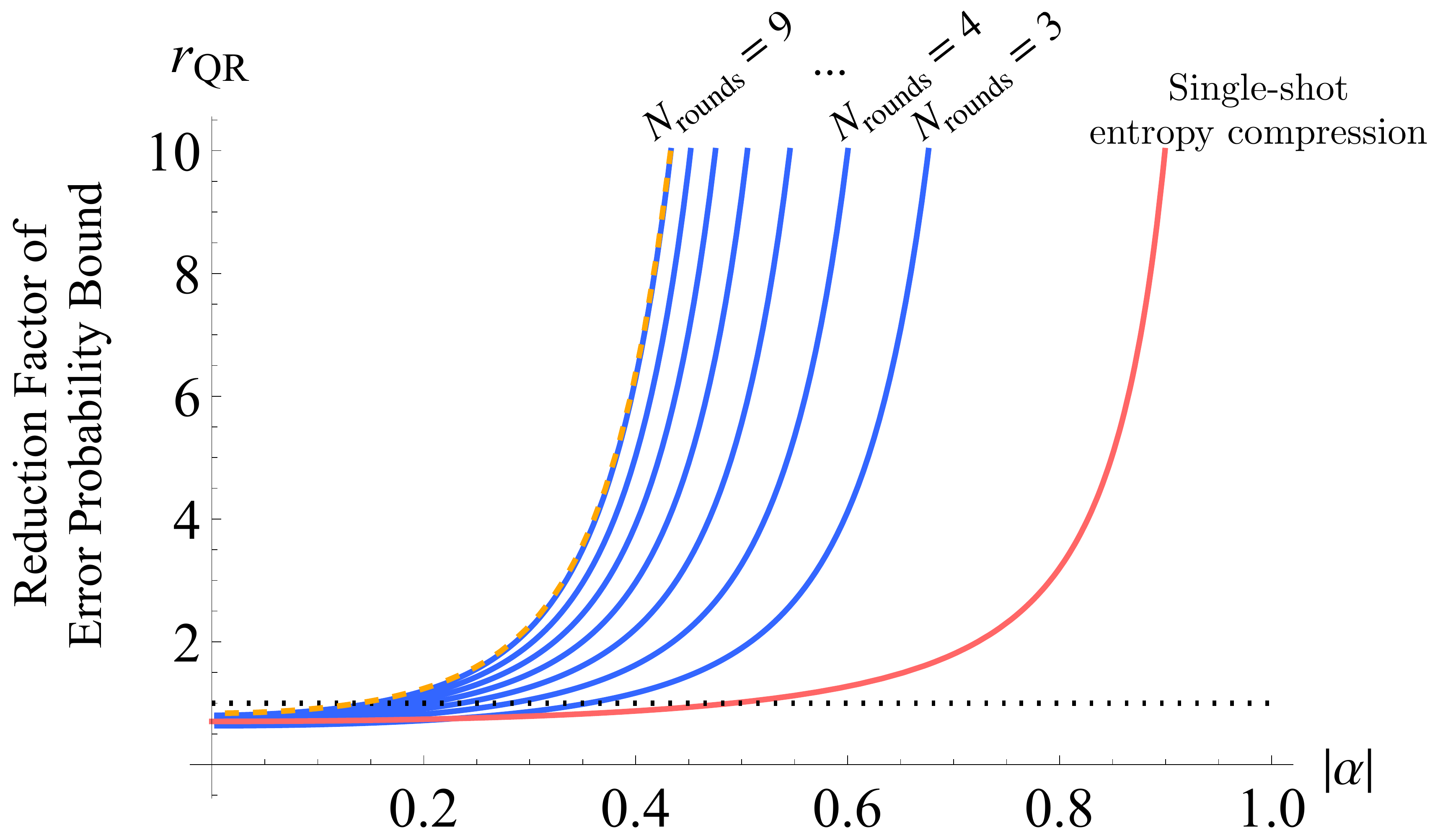}
\caption{Reduction factor of the error probability bound for the quantum refrigerator operating with $n=5$ qubits, shown in blue for different numbers of rounds as a function of the initial $|\alpha|$. The pink line represents the improvement achieved through single-shot entropy compression for $n=5$, while the yellow dashed line depicts the upper bound from simulations using optimal compressions for the case of $N_{\rm rounds}=9$. Notably, the performance of the BQR closely aligns with the optimal scenario, indicating an almost exact match for the configuration $n=5$, $m=2$, and $N_{\rm rounds}=9$.}
\label{fig_rqrn5}
\end{figure}
%

%(See supplemental material). To illustrate how the achievable polarization for the target qubit behaves, let's take the case of $n=4$:
%
%\begin{equation}
 %   \alpha_\textsc{qr}(n=4,N_{\rm rounds},\alpha)=
%\end{equation}
%
%{\color{pink} [It is possible to obtain the analytical results but they are very long and not so insightful... so maybe I will only describe it and use the numerical results]}

The enhanced polarization of the target qubit, $\alpha_\textsc{qr}$, depends on the refrigerator's configuration ($n$, $N_{\rm rounds}$, $m$) and the initial polarization of the sample qubits. Figure~\ref{fig:alphaQRn5} illustrates an example of this enhancement for $n=5$, showing results for several numbers of rounds as a function of the initial polarization $\alpha$. More details are provided in Appendix~\ref{sec:appendix_B}. 
The quantum refrigerator can achieve asymptotically the maximum polarization of the optimal HBAC~\cite{PhysRevLett.116.170501,lin2024thermodynamic} as the number of rounds increases: $\displaystyle\alpha_\infty=\frac{(1+\alpha)^{m2^{n-m-1}}-(1-\alpha)^{m2^{n-m-1}}}{(1+\alpha)^{m2^{n-m-1}}+(1-\alpha)^{m2^{n-m-1}}}$. However, it is unnecessary for the quantum refrigerator to operate near the cooling limit. Instead, the optimal configuration involves a small number of rounds, balancing significant polarization enhancement with efficient use of qubit resources to maximize the reduction of sampling error. The first few rounds provide the most substantial polarization enhancements, while subsequent rounds yield diminishing returns. The error reduction ratio, based on Eq.~(\ref{eq:error_predict}) and the enhanced polarization of the quantum refrigerator, is given by
\begin{align}
    \displaystyle
  r_\textsc{qr}= %&\frac{\rm Pr[error]}{\rm Pr_\textsc{qr}[error]}\\=&
  \frac{\alpha^{-2}-1}{\alpha_\textsc{qr}^{-2}\left(n,m,N_{\rm rounds},\alpha\right)-1}\cdot\frac{1}{(mN_{\rm rounds}+1)}
\label{eq:error_reduction_factor_qr}
\end{align}
where $\alpha_\textsc{qr}=\alpha_\textsc{qr}\left(n,m,N_{\rm rounds},\alpha\right)$ denotes the refrigerator's enhanced polarization.% after $N_{\rm rounds}$ with $m$ qubits refreshed per round. 

\begin{comment}
Figure~\ref{fig_rqrn5} illustrates the error probability reduction factor for the quantum refrigerator operating with $n=5$ qubits across various numbers of rounds, shown in blue. The pink line represents the improvement achieved through single-shot entropy compression, while the yellow dashed line indicates the upper bound derived from numerical simulations of the refrigerator using optimal compressions in each round for $N_{\rm rounds}=9$. Notably, in this configuration, the performance of the BQR closely aligns with the optimal scenario, demonstrating an almost exact match for $N_{\rm rounds}=9$.It is worth mentioning that for a smaller number of rounds, a larger gap exists between the optimal performance and the BQR, which is not depicted in this figure.
\end{comment}

Figure~\ref{fig_rqrn5} illustrates the error probability reduction factor for the quantum refrigerator operating with $n=5$ qubits, shown in blue for various numbers of rounds. The pink line represents the improvement achieved through single-shot entropy compression, while the yellow dashed line indicates the upper bound derived from numerical simulations using optimal compressions for the case of $N_{\rm rounds}=9$. In this configuration, the BQR's performance closely aligns with the optimal scenario, demonstrating an almost exact match for $N_{\rm rounds}=9$.
It is important to note that the upper bound depends on the number of rounds, with only the case of $N_{\rm rounds}=9$ displayed in the figure. For smaller numbers of rounds in this example ($N_{\rm rounds}=3$ to $8$), the upper bound is not depicted, but the gap between the BQR and the optimal bound is slightly larger. This gap gradually narrows as the number of rounds increases,  reaching its minimum at a relatively small number of rounds. As discussed earlier, achieving the optimal configuration does not require operating near the cooling limit or using an excessively large number of rounds. Instead, it necessitates balancing significant polarization enhancement with efficient use of qubit resources to minimize sampling error.

The BQR significantly enhances the sampling error probability bound, both on average across all initial polarizations for uniformly distributed $\alpha$ and locally within the high-polarization regime. Furthermore, although the protocol does not provide improvements in the low-polarization regime, the region where it offers an advantage can be extended by optimizing the number of qubits in the quantum refrigerator and the number of rounds.

\subsection{BQR with $k-$local Compressions}\label{sec:BQR_klocal}
\begin{figure*}[htbp]
%\begin{figure}[h]
\centering
\includegraphics[width=0.9\linewidth]{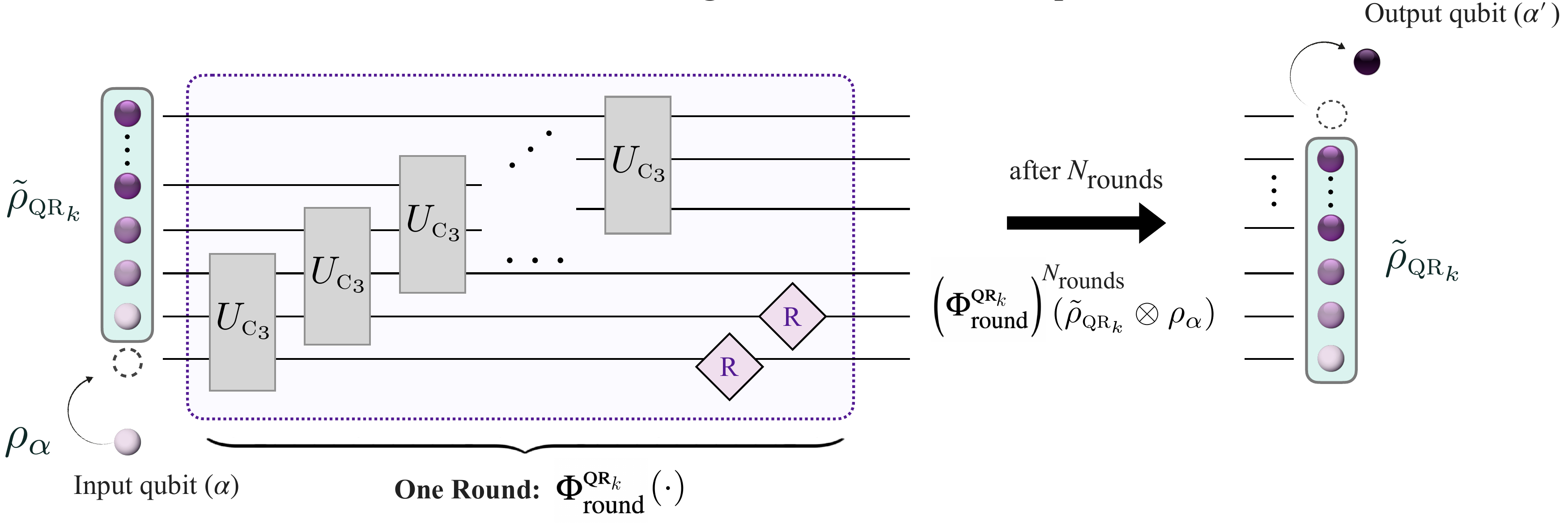}
\caption{Bidirectional quantum refrigerator protocol with $k$-local compression. The image illustrates the example for $k=3$. }
\label{fig:PracticalQR}
\end{figure*}

%\textit{Practical Bidirectional Quantum Refrigerator.---} 
In this section, we introduce a more practical version of the quantum refrigerator, in which the entropy compression step is performed using $k$-local unitaries. These unitaries are applied sequentially, acting on every set of $k$ neighboring qubits in a staircase-like manner, as illustrated in Fig.~\ref{fig:PracticalQR} for the case of $k=3$. The proposed protocol significantly enhances implementability while preserving the key advantages of the quantum refrigerator.

To illustrate their applicability and results, in this section, we focus our analysis on the $k=3$ scenario, with the $U_{C_3}$ as the $3$-local unitary, and using two reset qubits, as shown in Fig.~\ref{fig:PracticalQR}. The global unitary corresponding to a round of the refrigerator, $U_{{\rm QR}{(k=3)}}=(U_{C_3}\otimes\mathds{1}_{2^{n-3}})(\mathds{1}_{2}\otimes U_{C_3}\otimes\mathds{1}_{2^{n-4}})...(\mathds{1}_{2^{n-3}}\otimes\ U_{C_3})$,
%where $\mathds{1}_{j}$ represents the identity operator that acts on $j$ qubits, 
gives an asymptotic target polarization of $\alpha^\infty_n(k=3)=2 p_\alpha^{F_{n}}/\left( \sum_{i=0}^{F_{n}-1}(-1)^i\binom{F_{n}}{i}p_\alpha^i\right)-1$, where $F_n$ is the $n^{\rm th}$ Fibonacci number (i.e. $F_n=F_{n-1}+F_{n-2}$, with $F_1=1$ and $F_2=1$; see details in Appendix~\ref{sec:appendix_B}). However, as mentioned in previous section, it is not necessary for the quantum refrigerator to operate near the cooling limit. Instead, the optimal configuration involves a small number of rounds, balancing significant polarization enhancement with the efficient use of qubit resources to maximize the reduction in sampling error.

\begin{figure}[t]
\centering
\includegraphics[width=1.045\linewidth]{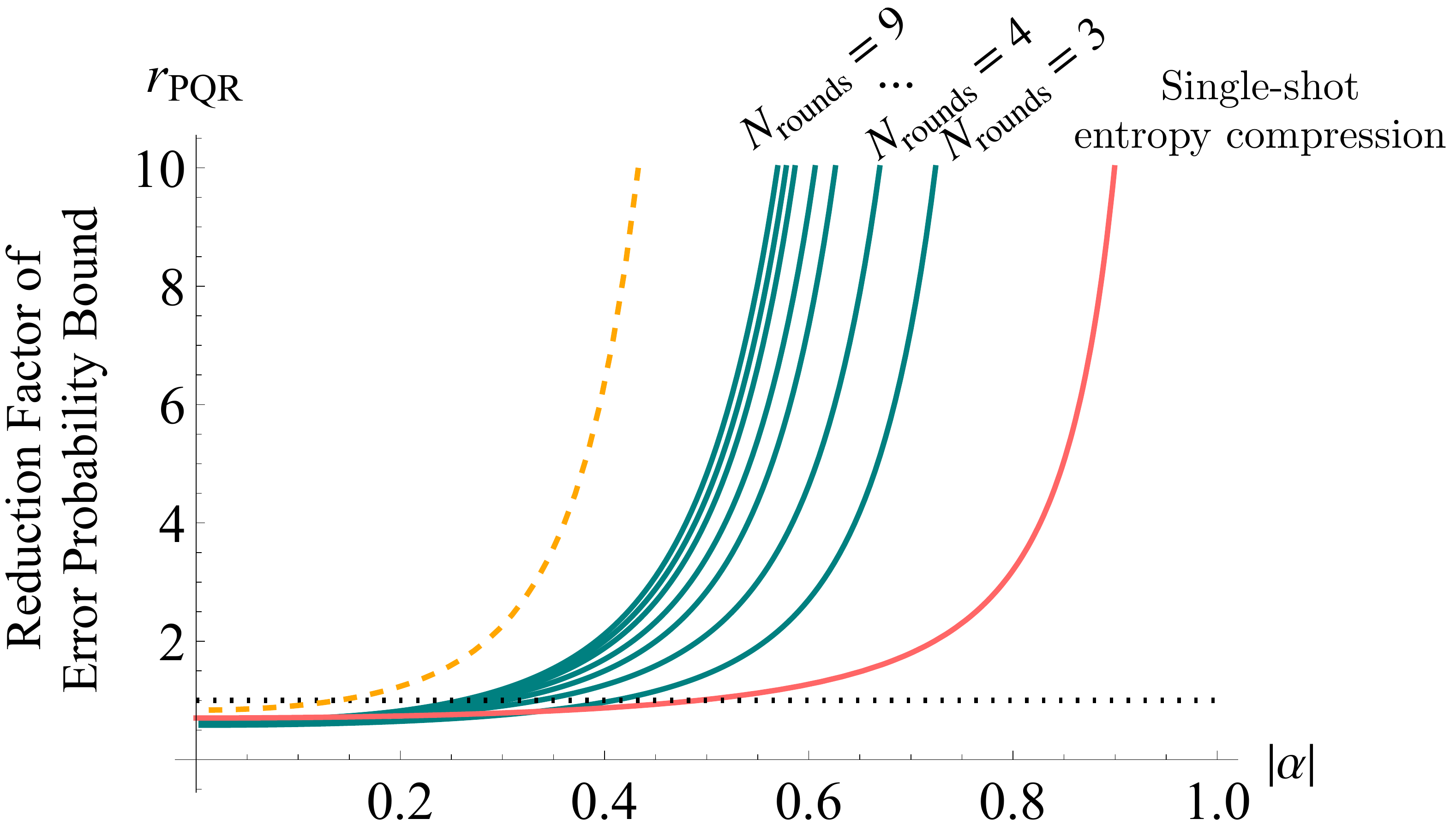}
\caption{Reduction factor of the error probability bound for the 3-local compression quantum refrigerator operating with $n=5$ qubits, shown in green for different numbers of rounds as a function of the initial $|\alpha|$. The pink line represents the improvement achieved through single-shot entropy compression for $n=5$, while the yellow dashed line indicates the upper bound from simulations with optimal compressions for $N_{\rm rounds}=9$. Although the performance of the 3-local approach shows a noticeable gap from the upper bound, reflecting reduced optimality, this method offers significantly greater practicality for implementation.}
\label{fig_practicalkkacm5q}
\end{figure}

The reduction factor of the error probability bound using the $k$-local compression, $r_{\textsc{qr}_{k{\rm -local}}}$ is given by Eq.~(\ref{eq:error_reduction_factor_qr}), with the enhanced polarization $\alpha_\textsc{qr}$ corresponding to the enhancement in the $k$-local compression case. 
Figure~\ref{fig_practicalkkacm5q} shows the reduction factor of the error probability bound for the 3-local compression quantum refrigerator operating with $n=5$ qubits for different numbers of rounds as a function of the initial $|\alpha|$. The pink line represents the improvement achieved through single-shot entropy compression for $n=5$, while the yellow dashed line indicates the upper bound from simulations with optimal compressions for $N_{\rm rounds}=9$. Note that even though the original quantum refrigerator achieves a greater improvement, the $k$-local compression version also enables a significant reduction in the error estimation bound, both on average when the polarization is uniformly distributed and outside the low-polarization regime. 

%
%
%
%{\color{pink} Here I will add the steady state of the Practical QR for a finite number of rounds. And the maximum reduction ratio as a function of $n$}

\begin{comment}
\subsection{Upper Bound for $r_\textsc{qr}$}

Although conventional HBAC is not directly implementable to assist in the classification problem, the HBAC optimal entropy compression can be useful for establishing a bound on the performance of BQRs, for the given configuration $\{n, N_{\rm rounds},m,\alpha\}$. Specifically, the operation of a refrigerator employing HBAC's optimal entropy compressions can be numerically simulated. In this approach, each compression involves performing a descending sort of the diagonal elements of the system's total density matrix (or an ascending sort for the opposite effect). This procedure establishes an upper bound on the performance achievable by the refrigerator.
%
In Figs.~\ref{fig_rqrn5} and~\ref{fig_practicalkkacm5q}, the yellow dashed line represents an example of the upper bound, shown in comparison to our Bidirectional Quantum Refrigerator (BQR) and the $3$-local version, respectively, for $n=5$ and $N_{\rm rounds}=9$. For the BQR, the yellow dashed line closely aligns with the bound, indicating an almost exact match. In contrast, the $3$-local case exhibits a noticeable gap, reflecting less optimal performance relative to the bound. Nonetheless, as discussed, the $3$-local approach offers greater practicality for implementation.

\end{comment}

\section{Conclusions and Discussion}
\label{sec:conc}

This study integrates quantum thermodynamics with QML to enhance sampling efficiency through a novel bidirectional quantum refrigerator technique. By conceptualizing quantum supervised learning as a thermodynamic cooling process, the proposed method significantly reduces the finite sampling error or the number of repetitions required for accurate classification, without the need for Grover-like operations. This practical approach applies to various QML models, including variational quantum circuits and quantum kernel methods, demonstrating its versatility and broad impact. The technique is particularly suited for NISQ devices, addressing the current limitations of quantum hardware and making QML algorithms more practical and scalable. 
% Furthermore, by utilizing single-qubit measurements, the method effectively lessens the barren plateau problem for shallow-depth circuits, enhancing the robustness and reliability of variational quantum algorithms. 
The interdisciplinary nature of this work, bridging quantum information processing, thermodynamics, and data science, not only advances the field of QML but also opens new research avenues in quantum thermodynamics, particularly for advancing algorithmic cooling techniques.

A remaining challenge is determining whether the proposed cooling protocol is optimal for reducing finite sampling errors. If optimality cannot be established, further refinement of the protocol to enhance its performance presents a compelling direction for future research. Additionally, investigating how coherence and non-classical correlations within the system and bath qubits can be harnessed to improve cooling efficiency offers an intriguing avenue. A detailed quantitative analysis of how our method mitigates the barren plateau effect also deserves further investigation. Another open question is whether these cooling techniques can be adapted to reduce finite sampling errors in the estimation of quantum kernels. This application presents unique challenges, as the quantities to be estimated involve the kernel matrix elements, rather than simple binary outcomes. Consequently, the naive application of the proposed method, which primarily aids in sign estimation, would not directly suffice. Addressing this challenge will broaden the applicability of insights from quantum thermodynamics to QML.

\section*{Acknowledgments}
This work was supported by Institute of Information \& communications Technology Planning \& evaluation (IITP) grant funded by the Korea government (No. 2019-0-00003, Research and Development of Core Technologies for Programming, Running, Implementing and Validating of Fault-Tolerant Quantum Computing System), the Yonsei University Research Fund of 2024 (2024-22-0147), and the National Research Foundation of Korea (Grant No. 2023M3K5A1094813).  N.A.R.B. acknowledges funding from the European Research Council (Consolidator Grant ‘Cocoquest’ 101043705) and the Austrian Research Promotion Agency (FFG) through the project FO999914030 (MUSIQ), funded by the European Union – NextGenerationEU. N.A.R.B. also acknowledges support from the Miller Institute for Basic Research in Science at the University of California, Berkeley, during the initial months of the project. We thank Hyukjoon Kwon for helpful discussions.

\bibliographystyle{unsrt}
\bibliography{references}

%%%%%%%%%%%%%%%%%%%%%%%%%%%%%%%%%%%%%%%%%%%%%%%%%%%%%%%%%%%
%%%%%%%%%%%%%%%%%%%%%%%%%%%%%%%%%%%%%%%%%%%%%%%%%%%%%%%%%%%

%\newpage
% \begin{center}
% {\large SUPPLEMENTAL MATERIAL}
% \end{center}
% \label{sm:supl_a}
% \vspace{4mm}

\appendix

\section{Optimal single entropy compression}

\subsection{Approximation of the reduction of finite sampling error}
\label{sec:appendix_A}

In this subsection, we present the approximations of the reduction in the sampling error probability bound after implementing an optimal entropy compression on $n$ qubits, in different polarization regimes. The expression of the factor of enhancement $r_\textsc{ac}(n,\alpha,\alpha_\textsc{ac})$ of Eq.~(\ref{eq:kkcom}), with enhanced polarization $\alpha_\textsc{ac}$, Eq.~(\ref{eq:alphaAC}), and the estimation error that satisfies the right sign of the expectation value ($\epsilon(\alpha)=|\alpha|$), can be rewritten as

\begin{equation}
\displaystyle r_\textsc{ac}(\alpha,\alpha_\textsc{ac},n)=\frac{1}{n}\cdot\frac{1-\alpha^{-2}}{1-\alpha_\textsc{ac}^{-2}}.
\label{eq:rac_app}
\end{equation}

To approximate that expression in the different approximation regimes, we have the following considerations:

$\displaystyle\alpha_\textsc{ac} = 2\left[\sum^{\lfloor(n-1)/2\rfloor}_{i=0}\binom{n}{i}p^{n-i}\left(1-p\right)^{i}\right]-1$, where $\displaystyle p=(1+\alpha)/2$, and $\lfloor\cdot\rfloor$ denotes the floor function. The enhanced polarization $\alpha_\textsc{ac}$ can be approximated to $\displaystyle \alpha_\textsc{ac}\simeq {\rm erf}(\xi):=\frac{2}{\sqrt{\pi}}\int^\xi_0 e^{-t^2}dt$, where $\displaystyle\xi=n\alpha/\sqrt{2n(1-\alpha^2)}$. Then, the reduction factor of probability error bound $r_\textsc{ac}$ can be rewritten as
\begin{equation}
        \displaystyle
        r_\textsc{ac}\simeq\frac{1}{n}\cdot\frac{1-\alpha^{-2}}{1-{\rm erf}\left(\frac{n\alpha}{\sqrt{2n\left(1-\alpha^2\right)}}\right)^{-2}}.
\end{equation}
\vspace{3pt}

\textbf{(1)} For small initial polarization $\alpha$: The argument of the error function can be approximated as $\displaystyle \xi\simeq\sqrt{\frac{n}{2}}\alpha$. Using the small-$\alpha$ expansion of $\displaystyle {\rm erf}(\xi)=\frac{2}{\sqrt{\pi}}\sum_{j=0}^{\infty}\frac{(-1)^j}{2j+1}\frac{\xi^{2j+1}}{j!}\simeq\frac{2}{\sqrt{\pi}}\xi=\frac{\sqrt{2n}\alpha}{\sqrt{\pi}}$. Substituting this into Eq.~(\ref{eq:rac_app}), the factor of enhancement simplifies as $\displaystyle r_\textsc{ac}\simeq\frac{2}{\pi}\cdot\frac{1-\alpha^2}{1-(2n\alpha^2/\pi)}\sim\frac{2}{\pi}$, in the small-polarization regime.

\vspace{3pt}

\textbf{(2)} In the intermediate polarization regime $\alpha$: Using the asymptotic expansion of the error function for the argument $\xi$:
$\displaystyle{\rm erf}\left(\xi\right)=1-\frac{e^{-\xi^2}}{\xi\sqrt{\pi}}\sum_{j=0}^{\infty}\left(-1\right)^j\frac{\left(2j-1\right)!!}{\left(2\xi^2\right)^j}$. For the intermediate polarization regime, this expression can be approximated to $\displaystyle{\rm erf}\left(\xi\right)\simeq1-\frac{e^{-\xi^2}}{\xi\sqrt{\pi}}$.
Then, the denominator of the reduction factor of the probability error bound $r_\textsc{ac}$ simplifies to $\displaystyle n\left( 1-{\rm erf}(\xi)^{-2}\right)\simeq n\left(1-\left(1+\frac{2e^{-\xi^2}}{\xi\sqrt{\pi}}\right)\right)=-\frac{2ne^{-\xi^2}}{\xi \sqrt{\pi}}$. And the full expression of $r_\textsc{ac}$ becomes
$\displaystyle r_\textsc{ac}\simeq\frac{-\xi\sqrt{\pi}e^{\xi^2}}{2n}(1-\alpha^{-2})$. Substituting $\xi=n\alpha/\sqrt{2n\left(1-\alpha^2\right)}$, the behavior of $r_\textsc{ac}$ is dominated by a term that grows exponentially as follows
\begin{equation}
    r_\textsc{ac}\propto e^{\xi^2}\;\; {\rm where}\;\;\xi^2=\frac{n\alpha^2}{2(1-\alpha^2)}.
\end{equation}
\vspace{3pt}

\textbf{(3)} For large initial polarization, $\alpha\sim1$: The denominator of $\displaystyle\xi=\frac{n\alpha}{\sqrt{2n(1-\alpha^2)}}$ approaches zero, causing the argument to diverge ($\xi\to\infty$). For large arguments ($\xi\to\infty$), ${\rm erf}(\xi)\to 1$. Thus, substituting this result into Eq.~(\ref{eq:rac_app}), the reduction factor diverges, $r_\textsc{ac} \to \infty$, in the large polarization regime.

\section{Evolution under the Bidirectional Quantum Refrigerator}
\label{sec:appendix_B}

The global effect of the set of unitaries in the Bidirectional Quantum Refrigerator (BQR) is given as follows:
\begin{align*}
    U_\textsc{qr}(n)&=U_{\textsc{c}_n}(\mathds{1}_{2}\otimes U_{\textsc{c}_{(n-1)}})(\mathds{1}_{2^2}\otimes U_{\textsc{c}_{(n-2)}})...(\mathds{1}_{2^{(n-3)}}\otimes U_{\textsc{c}_3}),
\end{align*}
where each $U_{\textsc{c}_j}$ has the effect of swapping the states $|0\rangle |1\rangle^{\otimes(j-1)}$ and $|1\rangle|0\rangle^{\otimes (j-1)}$. Specifically, each unitary $U_{\textsc{c}_j}$ has the following block matrix representation:
\begin{equation}
    U_{\textsc{c}_j}=\begin{bmatrix}
        \mathds{1}_{2^{(n-1)}-1} &0 &0\\
        0&\sigma_x &0\\
        0&0&\mathds{1}_{2^{(n-1)}-1}
    \end{bmatrix}
    \label{eq:Ucj}
\end{equation}

The general expression of $U_\textsc{qr}(n)$, can be obtained by replacing Eq.~(\ref{eq:Ucj}) as follows:

For $n=3$, 
\begin{equation}
    U_\textsc{qr}(3)=U_{\textsc{c}_3}=\begin{bmatrix}
        \mathds{1}_3 &0&0\\
        0&\sigma_x &0\\
        0&0&\mathds{1}_3
    \end{bmatrix},
\end{equation}
which corresponds to the optimal entropy compression of the 3-qubit HBAC method.

For $n=4$,
\begin{align}
    U_\textsc{qr}(4) &=U_{\textsc{c}_4}(\mathds{1}_2\otimes U_{\textsc{qr}}(3))=U_{\textsc{c}_4}(\mathds{1}_2\otimes U_{\textsc{c}_3})=\\ &=\begin{bmatrix}
        \mathds{1}_7 &0&0\\
        0&\sigma_x &0\\
        0&0&\mathds{1}_7
    \end{bmatrix}\left( \mathds{1}_2 \otimes \begin{bmatrix}
        \mathds{1}_3 &0&0\\
        0&\sigma_x &0\\
        0&0&\mathds{1}_3
    \end{bmatrix}\right)\\
    &=\begin{bmatrix}
        \mathds{1}_3 &0&0&0&0&0&0\\
        0&\sigma_x&0&0&0&0&0\\
        0&0 &\mathds{1}_2&0&0&0&0\\
        0&0&0&\sigma_x&0&0&0\\
        0&0&0&0&\mathds{1}_2&0&0\\
        0&0&0&0&0&\sigma_x&0\\
        0&0&0&0&0&0&\mathds{1}_3
    \end{bmatrix}=\\
    &=\begin{bmatrix}
        \mathds{1}_3 &0&0&0\\
        0&\mathds{1}_2\otimes\begin{bmatrix}
            \sigma_x&0\\
            0&\mathds{1}_2
        \end{bmatrix}&0&0\\
        0&0&\sigma_x&0\\
        0&0&0&\mathds{1}_3
    \end{bmatrix}.
\end{align}

Similarly, for general $n>3$, 
\begin{equation}
U_\textsc{qr}(n)=U_{\textsc{c}_n}(\mathds{1}_2\otimes U_\textsc{qr}(n-1)).    
\end{equation}
Thus, $U_\textsc{qr}(n)$ can be represented in matrix form as the following block matrix
\begin{align}
U_\textsc{qr}(n)=\begin{bmatrix}
\mathds{1}_3 & 0 &0\\
0 & \sigma_\textsc{qr}(n) & 0\\
0& 0 & \mathds{1}_3 
\end{bmatrix},
\end{align}
where $\sigma_\textsc{qr}(3)=\sigma_x$, and $\sigma_\textsc{qr}(n>3)$ is defined as
\begin{align}
        \sigma_\textsc{qr}(n):=\begin{bmatrix}
        \mathds{1}_2\otimes\begin{bmatrix} 
        \sigma_\textsc{qr}(n-1) &0\\
        0&\mathds{1}_2
        \end{bmatrix} &0\\
        0&\sigma_x
    \end{bmatrix},\;\;\;{\rm for} \;\;\; n>3,
\end{align}
or equivalently,
\begin{align*}
        \sigma_\textsc{qr}(n):=\begin{bmatrix}
        \mathds{1}_{(2^{n-2}-2)}\otimes\begin{bmatrix} 
        \sigma_x &0\\
        0&\mathds{1}_2
        \end{bmatrix} &0\\
        0&\sigma_x
    \end{bmatrix},\;\;\;{\rm for} \;\;\; n>3.
\end{align*}

\subsection{Evolution under the implementation of the BQR}

The effect of a single round on a system of $n$ qubits in the state $\rho$ is given by
\begin{equation}
    \rho\to \Phi^{\textsc{qr}}_{\rm round}(\rho):={\rm Tr_m}\left[ U_\textsc{qr}(n)\rho U^\dagger_\textsc{qr}(n)\right]\otimes\rho_\alpha^{\otimes m}.
    \label{eq:QRSingleRound}
\end{equation}
Then, the effect of the refrigerator operating with $N_{\rm rounds}$ is as follows:
\begin{equation}\rho\to \displaystyle
 \left(\Phi^{\textsc{qr}}_{\rm round}\right)^{N_{\rm rounds}}\left(\rho\right).
\end{equation}
The polarization of the target qubit is obtained as $\alpha'={\rm Tr}(Z\rho_{\rm target}^{\rm output})$, where $\displaystyle\rho_{\rm target}^{\rm output}={\rm Tr}_{\overline{\rm target}}
\left(\Phi^{\textsc{qr}}_{\rm round}\right)^{N_{\rm rounds}}\left(\rho\right)$ is the enhanced state of the target qubit after the implementation of the BQR.

Since the quantum refrigerator recycles the $n-1$ qubits remaining after extracting the target qubit and incorporates a fresh qubit from the sample, the updated input system is expressed as:
\begin{equation}
    \tilde{\rho}={\rm Tr_{target}}\left(\left(\Phi^{\textsc{qr}}_{\rm round}\right)^{N_{\rm rounds}}\left(\rho\right)\right)\otimes \rho_\alpha.
\end{equation}

After multiple iterations of the refrigerator, the state $\tilde{\rho}$ converges to a fixed point determined by the number of rounds, $N_{\rm rounds}$, and the number of reset qubits, $m$. In the asymptotic cooling limit, as $N_{\rm rounds}$ increases, the refrigerator reaches the maximum polarization achievable by the optimal  Heat-Bath Algorithmic Cooling (HBAC) protocol, as demonstrated in the next subsection. 

For the classification problem, however, it is not necessary to operate in the asymptotic cooling limit. Instead, the refrigerator will function at the fixed point corresponding to a smaller number of rounds $\tilde{\rho}_\textsc{qr}(n, N_{\rm rounds},m)$, as detailed in the section following the asymptotic discussion.

\subsubsection{Asymptotic polarization of the BQR}

After each round of the BQR, since $m$ qubits are reset, , the state of the system takes the form
\begin{equation}
    \rho\to\rho_{\rm comp}\otimes\rho_\alpha^{\otimes m}
\end{equation}
where $\rho_{\rm comp}$ represents the state of the compressed qubits after removing the reset qubits. Without loss of generality, the vector corresponding to the diagonal of $\rho_{\rm com}$ can be expressed as
\begin{equation}
    {\rm diag(\rho_{\rm comp})}=\begin{bmatrix}
        A_1\\A_2\\...\\A_{2^{n-m}}
    \end{bmatrix}.
\end{equation}
In particular, for the case $m=2$,  the form of the state is given as follows:
\begin{equation}
    {\rm diag}(\rho) \to \begin{bmatrix}
        A_1\\A_2\\...\\A_{2^{n-2}}
    \end{bmatrix}\otimes
    \begin{bmatrix}
        p_\alpha^2\\p_\alpha(1-p_\alpha)\\p_\alpha(1-p_\alpha)\\(1-p_\alpha)^2
    \end{bmatrix}
    \label{eq:diagrho}
\end{equation}
where $\displaystyle p_\alpha=\frac{1+\alpha}{2}$. Note that the elements of the density matrix for the reset qubits are already sorted in decreasing order when $\alpha>0$ or in increasing order when $\alpha<0$.

The full form of Eq.~(\ref{eq:diagrho}) is given as
\begin{equation}
    {\rm diag}(\rho)\to\begin{bmatrix}
        A_1 p_\alpha^2\\
        A_1 p_\alpha(1-p_\alpha)\\
        A_1 p_\alpha(1-p_\alpha)\\
        A_1 (1-p_\alpha)^2\\
        A_2 p_\alpha^2\\
        A_2 p_\alpha(1-p_\alpha)\\
        A_2 p_\alpha(1-p_\alpha)\\
        A_2 (1-p_\alpha)^2\\
        A_3 p_\alpha^2\\
        A_3 p_\alpha(1-p_\alpha)\\
        A_3 p_\alpha(1-p_\alpha)\\
        A_3 (1-p_\alpha)^2\\
        \cdot\\
        \cdot\\
        \cdot\\
        A_{2^{n-2}} p_\alpha^2\\
        A_{2^{n-2}} p_\alpha(1-p_\alpha)\\
        A_{2^{n-2}} p_\alpha(1-p_\alpha)\\
        A_{2^{n-2}} (1-p_\alpha)^2
           \end{bmatrix}
\end{equation}

After a new round of the refrigerator, the state is transformed in the following way,
\begin{equation}
   \begin{bmatrix}
        A_1 p_\alpha^2\\
        A_1 p_\alpha(1-p_\alpha)\\
        A_1 p_\alpha(1-p_\alpha)\\
       {\color{blue} A_1 (1-p_\alpha)^2}\\
       {\color{blue} A_2 p_\alpha^2}\\
        A_2 p_\alpha(1-p_\alpha)\\
        A_2 p_\alpha(1-p_\alpha)\\
      {\color{red}  A_2 (1-p_\alpha)^2}\\
       {\color{red} A_3 p_\alpha^2}\\
        A_3 p_\alpha(1-p_\alpha)\\
        A_3 p_\alpha(1-p_\alpha)\\
        {\color{magenta}A_3 (1-p_\alpha)^2}\\
        {\color{magenta} A_4 p_\alpha^2}\\
        \cdot\\
        \cdot\\
        \cdot\\
        {\color{cyan} A_{2^{n-2}-1}(1-p_\alpha)^2}\\
        {\color{cyan}A_{2^{n-2}} p_\alpha^2}\\
        A_{2^{n-2}} p_\alpha(1-p_\alpha)\\
        A_{2^{n-2}} p_\alpha(1-p_\alpha)\\
        A_{2^{n-2}} (1-p_\alpha)^2
           \end{bmatrix}
           \xrightarrow{\Phi^{\textsc{qr}}_{\rm round}}
              \begin{bmatrix}
              
        A_1 p_\alpha^2\\
        A_1 p_\alpha(1-p_\alpha)\\
        A_1 p_\alpha(1-p_\alpha)\\
       
        {\color{blue} A_2 p_\alpha^2}\\
        {\color{blue} A_1 (1-p_\alpha)^2}\\
        A_2 p_\alpha(1-p_\alpha)\\
        A_2 p_\alpha(1-p_\alpha)\\
       
        {\color{red} A_3 p_\alpha^2}\\
        {\color{red} A_2 (1-p_\alpha)^2}\\

        A_3 p_\alpha(1-p_\alpha)\\
        A_3 p_\alpha(1-p_\alpha)\\
        
        {\color{magenta} A_4 p_\alpha^2}\\
           {\color{magenta} A_3 (1-p_\alpha)^2}\\
        \cdot\\
        \cdot\\
        \cdot\\
         {\color{cyan} A_{2^{n-2}} p_\alpha^2}\\
        {\color{cyan} A_{2^{n-2}-1} (1-p_\alpha)^2}\\

        A_{2^{n-2}} p_\alpha(1-p_\alpha)\\
        A_{2^{n-2}} p_\alpha(1-p_\alpha)\\
        A_{2^{n-2}} (1-p_\alpha)^2
           \end{bmatrix},
\label{eq:vecOneRound}
\end{equation}
The elements in the diagonal that are permuted are the following:
\[
A_{(i+1)}p_\alpha^2 \xleftrightarrow{} A_i (1-p_\alpha)^2, \quad \text{for } i = 1, 2, \dots, 2^{n-2}-1,
\]
as shown in color in the transformation. 

In the asymptotic limit, as the number of rounds increases, the state of the elements converges to one that is invariant under the $\Phi^\textsc{qr}_{\rm round}$ operation. Consequently, in this asymptotic limit, the following condition must be satisfied: 
\begin{equation}
    A_{(i+1)}p_\alpha^2 = A_i (1-p_\alpha)^2, \quad \text{for } i = 1, 2, \dots, 2^{n-2}-1.
\end{equation}
This condition has the form of the one given by the conventional HBAC using 2 reset qubits, see Eq.~(S29) of the suplemental material in Ref.~\cite{PhysRevLett.116.170501}. From here, asymptotic polarization for the target qubit as the $N_{\rm rounds}$ grows is the maximum achievable for the HBAC with $m=2$:
\begin{align*}
\alpha_\infty(n,m=2)&=\frac{\left(1+\alpha\right)^{2^{n-2}}-\left(1-\alpha\right)^{2^{n-2}}}{\left(1+\alpha\right)^{2^{n-2}}+\left(1-\alpha\right)^{2^{n-2}}}\\
&={\rm tanh}\left[2^{n-2}{\rm arctanh}\left(\alpha\right)\right]
\end{align*}

Similarly, it is now straightforward to see that by following the same derivations, for any $m>2$, the condition that holds in the asymptotic cooling limit is given by:
\begin{equation}
        A_{(i+1)}p_\alpha^{m} = A_i (1-p_\alpha)^{m}, \quad \text{for } i = 1, 2, \dots, 2^{n-m}-1,
\end{equation}
which gives the general condition for the asymptotic cooling limit, see Eq.~(S29) of the suplemental material in Ref.~\cite{PhysRevLett.116.170501}: 
\begin{align}
\alpha_\infty(n,m)&=\frac{\left(1+\alpha\right)^{m2^{n-m-1}}-\left(1-\alpha\right)^{m2^{n-m-1}}}{\left(1+\alpha\right)^{m2^{n-m-1}}+\left(1-\alpha\right)^{m2^{n-m-1}}}\\
&={\rm tanh}\left[m2^{n-m-1}{\rm arctanh}\left(\alpha\right)\right]
\end{align}

\subsubsection{Enhanced polarization for the BQR operating with finite $N_{\rm rounds}$}

The enhanced polarization of the target qubit, $\alpha_\textsc{qr}$, depends on the refrigerator's configuration parameters ($n$, $N_{\rm rounds}$, and $m$) that determine the steady state of the BQR, $\tilde{\rho}_\textsc{qr}\left(n,N_{\rm rounds},\alpha\right)$. 

In this subsection, we present the general evolution under $N_{\rm rounds}$, the corresponding steady state of the refrigerator, and the enhanced polarization achieved for a general configuration. We then illustrate explicit examples of the enhancement for refrigerators operating with $n=4$ and $n=5$, both with $m=2$, as a function of the number of rounds.

In general, the state of the $n$ system immediately after a round,  specifically after the $j^{\rm th}$ round, takes the following form,
\begin{equation}
   {\rm diag} (\rho^{(j)})=\begin{bmatrix}
    A_1^{(j)}\\
    A_2^{(j)}\\
    ...\\
    A_{2^{n-m}}^{(j)}
\end{bmatrix}\otimes{\rm diag}(\rho_\alpha^m).
\end{equation}

Let us denote by $\vec{A}^{(j)}$ the diagonal vector of the state of the first $n-m$ qubits. By implementing one round of the BQR, following the transformation given in Eq.~(\ref{eq:QRSingleRound}), the diagonal vector is transformed using a stochastic matrix $M$ as follows:
\begin{equation}
    \vec{A^{(j)}} \xrightarrow{\Phi^{\textsc{qr}}_{\rm round}} \vec{A}^{(j+1)}=M\cdot \vec{A}^{(j)}.
    \label{eq:MatrixAvector}
\end{equation}
After $N_{\rm rounds}$, that diagonal vector is given by:
\begin{equation}
    \vec{A}^{N_{\rm rounds}}=M^{N_{\rm rounds}}\cdot \vec{A}^{(0)},
\end{equation}
where $\vec{A}^{(0)}$ represents to the initial state when the refrigerator begins operation. Since this state  is recycled from the previous cycle of the refrigerator, and $M^{N_{\rm round}}$ has attractive fixed points, the refrigerator will eventually operate in a steady state. In this regime, the input state to the refrigerator matches the output state after removing the enhanced target qubit. This condition implies that the following equation must be solved to determine the state $\tilde{\rho}_\textsc{qr}$:
\begin{equation}
    \tilde{\rho}_\textsc{qr}={\rm Tr_{target}}\left[\left(\Phi^{\textsc{qr}}_{\rm round}\right)^{N_{\rm rounds}}\left(\tilde{\rho}_\textsc{qr}\otimes\rho_\alpha\right)\right]
\end{equation}
Expressed in terms of the diagonal vector of the fixed point, $\vec{A}{\tilde{\textsc{qr}}}$, this condition becomes:
\begin{equation}
    \vec{A}_{\tilde{\textsc{qr}}}={\rm Tr_{target}}\left[M^{N_{\rm rounds}}\cdot \vec{A}_{\tilde{\textsc{qr}}}\right]\otimes{\rm diag}(\rho_\alpha),
\label{eq:conditionA}
\end{equation}
which must be solved for a given $N_{\rm rounds}$. 
The updated polarization $\alpha'$ for the target qubit, after the $N_{\rm rounds}$ is given by:
\begin{align}
    &\alpha'={\rm Tr}\left[Z\rho_{\rm target}\right], \;\; \\{\text{where}} \;\; &\rho_{\rm {\rm target}}={\rm Tr_{/\rm target}}\left[M^{N_{\rm rounds}}\cdot\vec{A}_{\tilde{\textsc{qr}}}\right]
    \label{eq:enhancedalphaA}
\end{align}

\begin{itemize}
    \item \underline{Example for $n=4$, with $m=2$ for $N_{\rm rounds}=3$}:
\end{itemize}

The state of the $n=4$ system immediately after a round that resets $m=2$ qubits, takes the form,
\begin{equation}
    {\rm diag}(\rho_4^{(j)})=\begin{bmatrix}
    A_1^{(j)}\\
    A_2^{(j)}\\
    A_3^{(j)}\\
    A_4^{(j)}
\end{bmatrix}\otimes{\rm diag}(\rho_\alpha^2).
\end{equation}
By implementing one round of the BQR, following the transformation given in Eq.~(\ref{eq:QRSingleRound}), the stochastic matrix $M$ that describes that transformation is
\begin{equation}
    \vec{A^{(j)}} \xrightarrow{\Phi^{\textsc{qr}}_{\rm round}} \vec{A}^{(j+1)}=M\cdot \vec{A}^{(j)},
\end{equation}
where
\begin{equation*}
    M_4=\begin{bmatrix}
        p_\alpha\left(2-p_\alpha\right) &p_\alpha^2 & 0&0\\
        \left(1-p_\alpha\right)^2 &2p_\alpha(1-p_\alpha) &p_\alpha^2 &0\\
        0 & \left(1-p_\alpha\right)^2 & 2p_\alpha\left(1-p_\alpha\right) &p_\alpha^2\\
        0 & 0& (1-p_\alpha)^2 & 1-p_\alpha^2
        \end{bmatrix}
\end{equation*}
This is derived as in Eq.~(\ref{eq:vecOneRound}), followed by tracing out the two reset qubits to obtain the updated diagonal vector $\vec{A}^{(j+1)}$.

Note that for $n=4$ and $m=2$, the vector $\vec{A}_{\tilde{\textsc{qr}}}$ is given in the form
$\vec{A}_{\tilde{\textsc{qr}}}=\{p_\textsc{qr},1-p_\textsc{qr}\}\otimes\{p_\alpha,1-p_\alpha\}$, since two qubits are reset and the target qubit is removed. 
Thus, we only need to solve the parameter $p_\textsc{qr}$.
Then, the condition given in Eq.~(\ref{eq:conditionA}) to obtain the state of the refrigerator, for this case when the refigerator is operating with $N_{\rm rounds}=3$ is expressed as
\begin{equation*}  p_\textsc{qr}=p_\alpha^2\cdot\frac{1+2p_\alpha+16p_\alpha^2-70p_\alpha^3+86 p_\alpha^4-34p_\alpha^5}{1-20p_\alpha^2+106p_\alpha^3-219p_\alpha^4+198p_\alpha^5-66p_\alpha^6}.
\end{equation*}
Substituting this result in Eq.~(\ref{eq:enhancedalphaA}) with the updated $\vec{A}_{\tilde{\textsc{qr}}}$, the enhanced polarization of the target qubis is given by:
\begin{align*}
    &\frac{\alpha'}{\alpha}=\\&\frac{943 - 1608 \alpha^2 + 2711 \alpha^4 - 2248 \alpha^6 + 
 925 \alpha^8 - 240 \alpha^{10} + 29 \alpha^{12}}{16\left(25 - 19 \alpha^2 + 59 \alpha^4 - 33 \alpha^6\right)}
\end{align*}

Note that the plots in the main manuscript were obtained through direct simulation of the circuit for the bidirectional refrigerator. Here, we provide examples to illustrate how exact analytical expressions can be derived.

\begin{itemize}
    \item \underline{Example for $n=5$ with $m=2$}:
\end{itemize}

The diagonal state of the $n=5$ system immediately after a round that resets $m=2$ qubits takes the form ${\rm diag}(\rho^{(j)})={A_1^{(j)},A_2^{(j)},...,A_8^{(j)}}\otimes{\rm diag}(\rho_\alpha)$. The corresponding stochastic matrix that transforms the diagonal vector $\vec{A}$, following the effect of a round $\Phi^{\textsc{qr}}_{\rm round}(\cdot)$, is given by
\begin{widetext}
\begin{equation}
    M_5=\begin{bmatrix}
         p_\alpha\left(2-p_\alpha\right) &p_\alpha^2 & 0&0&0&0&0&0\\
        \left(1-p_\alpha\right)^2 &2p_\alpha(1-p_\alpha) &p_\alpha^2 &0&0&0&0&0\\
        0 & \left(1-p_\alpha\right)^2 & 2p_\alpha\left(1-p_\alpha\right) &p_\alpha^2&0&0&0&0\\               
        0& 0 & \left(1-p_\alpha\right)^2 & 2p_\alpha\left(1-p_\alpha\right) &p_\alpha^2&0&0&0\\
        0& 0 &0& \left(1-p_\alpha\right)^2 & 2p_\alpha\left(1-p_\alpha\right) &p_\alpha^2&0&0\\
        0& 0 &0& 0&\left(1-p_\alpha\right)^2 & 2p_\alpha\left(1-p_\alpha\right) &p_\alpha^2&0\\
        0 & 0&0&0&0&0& (1-p_\alpha)^2 & 1-p_\alpha^2
    \end{bmatrix},
\end{equation}
\end{widetext}
and similarly for $n>5$, with $m=2$.

As a concrete example, solving for the case of $N_{\rm rounds}=3$ with the refrigerator running in the steady state yields target qubits with the following enhanced polarization:
$\alpha'=\alpha(1444856 - 3513097 \alpha^2 + 
     7571347\alpha^4 - 7884719\alpha^6 + 
     5496725\alpha^8 - 1352250 \alpha^{10} + 
     681150 \alpha^{12} - 838878 \alpha^{14} + 
     272130 \alpha^{16} - 39085 \alpha^{18} - 
     4929 \alpha^{20} + 1661 \alpha^{22} + 
     97 \alpha^{24})/(32 (21456 - 47629 \alpha^2 + 
     134063 \alpha^4 - 129925 \alpha^6 + 
     76115\alpha^8 + 11917\alpha^{10} + 
     5277\alpha^{12} - 14023 \alpha^{14} + 81 \alpha^{16} + 
     12\alpha^{18}))$. This expression provides the exact analytical form of the red line in Fig.~\ref{fig:alphaQRn5}, which corresponds to $\alpha_\textsc{qr}$ for $N_{\rm rounds} = 3$. Substituting this result into Eq.~(\ref{eq:error_reduction_factor_qr})  generates the corresponding line in Fig.~\ref{fig_rqrn5}. The figures were reproduced numerically by simulating the circuit. These sections are intended to illustrate how to derive exact analytical results.

%%%%%%%%%%%%%%%%%%%%
%%%%%%%%%%%%%%%%%%%%

\subsection{Evolution under the Implementation of the BQR with \textit{k}-local Compressions}

For the BQR with $k$-local compressions, the evolution follows the same form as given in Eq.~(\ref{eq:QRSingleRound}), but with a different unitary. In the example presented in the main manuscript, where $k=3$, the unitary for each round is given by:
\begin{equation*}
U_{{\rm QR}{(k=3)}} = 
(U_{C_3} \otimes \mathds{1}_{2^{n-3}})
(\mathds{1}_{2} \otimes U_{C_3} \otimes \mathds{1}_{2^{n-4}}) \cdots 
(\mathds{1}_{2^{n-3}} \otimes U_{C_3}).
\end{equation*}
Thus, the effect of a single round on a system of $n$ qubits in the state $\rho$ is given by
\begin{equation}
     \Phi^{\textsc{qr}(k=3)}_{\rm round}(\rho) := 
     {\rm Tr}_m \left[ 
     U_{{\rm QR}{(k=3)}} \rho U^\dagger_{{\rm QR}{(k=3)}}
     \right] \otimes \rho_\alpha^{\otimes m}.
\end{equation}

A similar analysis for the steady state of the quantum refrigerator and the enhanced polarization applies in this case. The steady-state arises from the unitary $U_{{\rm QR}{(k=3)}}$ per round, instead of the $U_\textsc{qr}$.\\

\subsubsection{Asymptotic polarization of the BQR with \textit{3}-local compressions}

\begin{itemize}
    \item \underline{Asymptotic polarization for $n=4$ and $m=2$}:    
\end{itemize}

The stochastic matrix that describes the evolution in the form of Eq.~(\ref{eq:MatrixAvector}) is given by
\begin{equation*}
    M_{4(3{\rm -local})}=\begin{bmatrix}
        p_\alpha\left(2-p_\alpha\right) &p_\alpha^2 & 0&0\\
        \left(1-p_\alpha\right)^2 &p_\alpha(1-p_\alpha) &p_\alpha &0\\
        0 & 1-p_\alpha & p_\alpha\left(1-p_\alpha\right) &p_\alpha^2\\
        0 & 0& (1-p_\alpha)^2 & 1-p_\alpha^2
        \end{bmatrix}.
\end{equation*}

In the asymptotic limit, as the number of rounds increases, the vector $\vec{A}_\infty$ is invariant under the effect of one more round:
\begin{equation}
    \vec{A}_\infty=M_{4(3\rm -local)}\cdot\vec{A}_\infty.
\end{equation}
Solving this condition, we found that, in the asymptotic limit, the $n$ qubits are in a product state, with each qubit having a different ground state population. Considering the qubits in order, starting from the end of the string and moving towards the target qubit, the populations are given by
\begin{equation*}
\vec{p}^\infty(n=4)=\{p_\alpha,p_\alpha,\frac{p_\alpha^2}{1-2p_\alpha+2p_\alpha^2},\frac{p_\alpha^3}{1-3p_\alpha+3p_\alpha^2}\}.
\end{equation*}

The two qubits at the end of the string retain their initial populations, $p_\alpha$, as they are reset after each round. In contrast, the other qubits in the string exhibit improved populations.

As the size of the string increases, the steady state follows a specific pattern, which will be described in the following subsections.

\begin{itemize}
    \item \underline{Asymptotic polarization for $n=5$ and $m=2$}    
\end{itemize}

The stochastic matrix that describes the evolution in the form of Eq.~(\ref{eq:MatrixAvector}) is given by
\begin{widetext}
    
\begin{equation*}
    M_{5(3-{\rm local})}=\begin{bmatrix}
p_\alpha (2 - p_\alpha) & p_\alpha^2 & 0 & 0 & 0 & 0 & 0 & 0 \\
(1 - p_\alpha)^2 & p_\alpha (1 - p_\alpha) & p_\alpha & 0 & 0 & 0 & 0 & 0 \\
0 & 1 - p_\alpha & p_\alpha (1 - p_\alpha) & p_\alpha^2 & 0 & 0 & 0 & 0 \\
0 & 0 & 0 & 0 & p_\alpha (2 - p_\alpha) & p_\alpha^2 & 0 & 0 \\
0 & 0 & (1 - p_\alpha)^2 & 1 - p_\alpha^2 & 0 & 0 & 0 & 0 \\
0 & 0 & 0 & 0 & (1 - p_\alpha)^2 & p_\alpha (1 - p_\alpha) & p_\alpha & 0 \\
0 & 0 & 0 & 0 & 0 & 1 - p_\alpha & p_\alpha (1 - p_\alpha) & p_\alpha^2 \\
0 & 0 & 0 & 0 & 0 & 0 & (1 - p_\alpha)^2 & 1 - p_\alpha^2
\end{bmatrix}.
\end{equation*}
\end{widetext}

Solving the condition $\vec{A}_\infty=M_{5(3\text{-local})}\cdot\vec{A}_\infty.$, the corresponding vector of enhanced ground state populations of each qubit is given as
\begin{align*}
\vec{p}_\infty(n)=\{&p_\alpha,p_\alpha,\frac{p_\alpha^2}{1-2p_\alpha+2p_\alpha^2},\frac{p_\alpha^3}{1-3p_\alpha+3p_\alpha^2},\\
&\frac{p_\alpha^5}{1-5p_\alpha+10p_\alpha^2-10 p_\alpha^3+5p_\alpha^5}\}.
\end{align*}

\begin{itemize}
    \item \underline{Asymptotic polarization for $n$ and $m=2$ for the BQR} \underline{with $3$-local compressions}    
\end{itemize}

By iterating the described derivations, in the asymptotic limit of the BQR with $3$-local compressions the $n$ qubits will be in a product state, with the following list of the local ground state populations for the qubits, counting from the end of the string of qubits to the target qubit:
\begin{align*}
    \displaystyle
    \vec{p}_\infty(n)=\{p_\alpha\;,p_\alpha\;,\frac{p_\alpha^2}{1-2p_\alpha+2p_\alpha^2}\;,\;
    \frac{p_\alpha^{F_4}}{\displaystyle \sum_{i=0}^{F_4-1}(-1)^i\binom{F_4}{i}p_\alpha^i}\;, \; \\\frac{p_\alpha^{F_5}}{\displaystyle \sum_{i=0}^{F_5-1}(-1)^i\binom{F_5}{i}p_\alpha^i}\;,...,\;\frac{p_\alpha^{F_{n}}}{\displaystyle \sum_{i=0}^{F_{n}-1}(-1)^i\binom{F_{n}}{i}p_\alpha^i}\}\;\;{\rm for}\;\;n\geq3
\end{align*}
where $F_j$ is the $j^{\rm th}$ Fibonacci number (i.e. $F_j=F_{j-1}+F_{j-2}$ with $F_1=1$, $F_2=1$). 

However, as mentioned in the main text, it is not necessary for the quantum refrigerator to operate near the asymptotic cooling limit. Instead, the optimal configuration requires only a small number of rounds, striking a balance between significant polarization enhancement and efficient use of qubit resources to maximize the reduction in sampling error, as illustrated in Fig.~\ref{fig_practicalkkacm5q}.\\

\section{Estimating the gradient of the classification score}
\label{sec:appendix_C}

Writing the classification score as $q(x,\theta) = \langle x | \exp(i\theta G) M \exp(-i\theta G) |x\rangle$ with a Hermitian operator $G$, the gradient can be expressed as
\begin{equation}
\label{eq:q_grad1}
    \frac{\partial q(x,\theta)}{\partial \theta} = \langle x | \exp(i\theta G) i[G,M] \exp(-i\theta G)|x\rangle.
\end{equation}
Since $i[G,M]$ is Hermitian, the above equation represents measuring the expectation value of $\tilde{M} := i[G,M]$ on the same VQC that is used to compute $q(x,\theta)$.

In our binary classification setting, $M$ is assumed to be a Pauli observable. Therefore, if $G$ is a Pauli operator, then $\tilde{M}$ is also a Pauli operator. In this case, the Clifford transformation technique allows us to compute the gradient as
\begin{equation}
\label{eq:q_grad2}
    \frac{\partial q(x,\theta)}{\partial \theta} = \mathrm{Tr}\left(Z\sigma_1(x,\theta)\right),
\end{equation}
where $\sigma_1(x,\theta) = \mathrm{Tr}_{n-1}\left( V_c\exp(-i\theta G)|x\rangle \langle x | \exp(i\theta G) V^{\dagger}_c\right)$ and $V_c$ represents a Clifford gate. This expression is analogous to Eq.~(\ref{eq:score}) of the main text, except for the difference between $U_c$ and $V_c$, as $U_c$ transforms $M$ to $Z_1$ and $V_c$ transforms $\tilde{M}$ to $Z_1$. Thus, estimating the gradient with an additive error less than its magnitude results in a similar error probability bound as described in Eq.~(\ref{eq:error_predict}) of the main text.

\section{Kernel-based quantum binary classifiers}
\label{sec:appendix_D}
Although the main text focuses primarily on variational quantum binary classifiers (VQBCs), the transformation in Eq.(\ref{eq:goal}) can be applied more broadly to any binary classifier established based on the classification score in Eq.~(\ref{eq:score}). For instance, quantum kernelized binary classifiers (QKBC) utilizing Hadamard test or swap test circuits~\cite{blank2020quantum,park2021robust,QML_Maria_Francesco,PARK2020126422,Blank_2022,DEOLIVEIRA2024127356,LeeParkAQT2024} also compute the classification score as expressed in Eq.~(\ref{eq:score}). In such instances, the quantum circuits are designed so that the single-qubit Pauli-$Z$ measurement yields
\begin{equation}
\langle Z\rangle = \frac{1}{s}\sum_{i=1}^{s}y_ik(x_i,\tilde{x}),
\end{equation}
where $k(x_i,\tilde{x})$ represents the kernel function quantifying the similarity between a sample data point $x_i$ and the test data $\tilde{x}$, $y_m\in\lbrace +1,-1\rbrace$ denotes the class label for $x_i$, and $s$ is the total number of samples. The QKBC adheres to the same classification rule as described in Eq.~(\ref{eq:rule}) for the VQBC. 
Another notable example is the quantum support vector machine proposed in Ref.~\cite{PhysRevLett.113.130503_QSVM}. Therefore, the importance of developing an efficient technique to achieve Eq.~(\ref{eq:goal}) extends beyond VQBC, encompassing broader applications in QML.
\end{document}